\documentclass[preprint,rmp,aps]{revtex4}

\usepackage{graphics}
\usepackage{epsfig}

\begin{document}

\title{ Stochastic Dynamical Structure (SDS) of Nonequilibrium Processes in the Absence of Detailed Balance. IV: Emerging of Stochastic Dynamical Equalities and Steady State Thermodynamics
from Darwinian Dynamics}

\author{ P. Ao }

\address{ Department of Mechanical Engineering and Department of Physics,
          University of Washington, Seattle, WA 98195, USA }

\date{June 1, 2008 }

\begin{abstract}
The evolutionary dynamics first conceived by Darwin and Wallace,
referring to as Darwinian dynamics in the present paper, has been
found to be universally valid in biology.
The statistical mechanics and thermodynamics, while enormously
successful in physics, have been in an awkward situation of
wanting a consistent dynamical understanding. Here we present from
a formal point of view an exploration of the connection between 
thermodynamics and Darwinian dynamics and a few related topics.
We first show that the stochasticity in
Darwinian dynamics implies the existence temperature, hence the
canonical distribution of Boltzmann-Gibbs type. In term of relative
entropy the Second Law of thermodynamics is dynamically
demonstrated without detailed balance condition, and is valid
regardless of size of the system. In particular, the dynamical
component responsible for breaking detailed balance condition does
not contribute to the change of the relative entropy. Two types of
stochastic dynamical equalities of current interest are explicitly
discussed in the present approach: One is based on Feynman-Kac
formula and another is a generalization of Einstein relation. Both
are directly accessible to experimental tests. Our demonstration
indicates that Darwinian dynamics represents logically a simple
and straightforward starting point for statistical mechanics and
thermodynamics and is complementary to and consistent with
conservative dynamics that dominates the physical sciences.
Present exploration suggests the existence of a unified stochastic
dynamical framework both near and far from equilibrium. \\
{\it Emerging of Stochastic Dynamical Equalities and Steady State Thermodynamics
         from  Darwinian Dynamics},
        P. Ao, Communications in Theoretical Physics {\bf 49} (2008) 1073-1090.
           http://ctp.itp.ac.cn/qikan/Epaper/zhaiyao.asp?bsid=2817 ; \\ 
           http://www.iop.org/EJ/abstract/0253-6102/49/5/01 ; \\
           doi: 10.1088/0253-6102/49/5/01  \\
%
%  {\it This manuscript is to be published in Communications in Theoretical
%  Physics  (2008).} 
%
% \\
%05.70.Ln
% Nonequilibrium and irreversible thermodynamics \\
%05.10.Gg;
% stochastic analysis methods   (Fokker-Planck, Langevin, etc);
% \\n phenomena, random processes, noise,
%%     and Brownian motion
%% 87.80.Vt Dynamical, regulatory, and integrative biology
%72.70.+m;
% Noise processes and phenomena;  \\
%87.15.Ya
% Fluctuations

\end{abstract}

%\pacs{ PACS numbers:  ??? }

\maketitle

\tableofcontents

{\ }

{\ }

{\it One of the principle objects of theoretical research in any
department of knowledge is to find the point of view from which
the subject appears in its greatest simplicity.}

{ {\ }  \hspace{80mm} {\ } Josiah Willard Gibbs (1839-1903) }

%
%  J.W. Gibbs was the 5th PhD in USA, in 1863.
%
% In his letter of acceptance of the Rumford Medal in 1881.
% Cited in Mem of Physics: J. Willard Gibss, American mathematical physicist
% par excellence. R.J. Seeger. Pergamon, Oxford, 1974. pp 190.
%
% There are two versions:
%  "One of the principle objects of theoretical research in    my
% department of knowledge is to find the point of view from which
% the subject appears in its greatest simplicity."
%   in A.L. Mackay, Dictionary of Scientific Quotations. (London, 1994)
%
%  "One of the principle objects of theoretical research
% is to find the point of view from which
% the subject appears in its greatest simplicity."
%   heading of Gibbs bust at New York.
%
% Some other Gibbs' aphorisms, or attributed to him:
%  "The whole is simpler than its parts."
%  "In mathematics, a part often contains the whole."
%         1888. (collected works. pp115, v.2).
%  "The more we study the subject, the more we find all that is most useful
% and beautiful attaching itself to a few central principles."
%         1888.(collected works. pp115, v.2).
%  "Mathematics is a language."  recalled by G.N. Lewis.
%
% structures of theories
%   economy, efficiency, useful
% axiomatism: looking for a minimum structure
%   it is a great instrument for teaching.
% two falsifiers for physical theories:
%   consistency (with known experience) and experiment (with new experience)
%

\section{Introduction}

\subsection{What is Darwinian dynamics?}

The dynamical theory proposed by Darwinian and Wallace 
%(Darwin and Wallace, 1858; Darwin, 1958)
\cite{darwin1858,darwin1958}
on evolution in biology has formed the fundamental theoretical
structure for understanding biological phenomena for one
and half centuries. We refer to it in the present paper as
"Darwinian dynamics". It has been confirmed by thousands and
thousands of field observations and laboratory experiments, and
extended to virtually all levels of biology. There is no known
valid empirical evidence against it in biology, as in the same status of
relativity and quantum mechanics in physics. There exists a
concise and accessible discourse of this dynamics by a renowned
researcher in physical sciences that we recommend to the reader
\cite{schuster2005}. Its essence may be summarized by a single word
equation most familiar to biological scientists \cite{mayr2004}:
%
%R. Dawkins, Darwin Triumphant: Darwinism as a universal truth
%  in Man and Beast: revisited, 1991.
% Core Darwinism ... is the minimal theory that evolution is guided
% in adaptively nonrandom directions by nonrandom survival of
% small random hereditary changes. pp 26.
% Attributed to Williams (1966): There is no reason to assume that all
% evolutionary change is adaptive.
%  Does this imply the antisymmetric matrix?
% or merely in the fluctuation-dissipation theorem sense?
% or in the Wright's shift balance sense?
%
%
% Related issues in de-coherence theory in quantum mechanics
% Good side:
% The omnipresent interactions of the system with the degrees
% of freedom of its environment must be taken into account.
% Problems:
% 1) Decoherence alone cannot solve the measurement problem;
% 2) The requirement for a division of the universe into "systems"
% and "environments" introduces a strong flavor of subjectivity,
% since no general and objective rule exists for how and where to place the cut;
% 3) The necessity for an "external" environment leads to difficulties
% when one would like to apply the theory to the universe as a whole,
% as in quantum cosmology.
%  M. Schloshauer and A. Fine, in Qauntum Mechanics in a crossroad (2007)
%
% Four elements of decohernece theory:
% von Neumann's projectors; decoherence;
% derirvation of classical physics; consistent histories
%  R. Omnes, in Qauntum Mechanics in a crossroad (2007)
%

\begin{center}
  {\it Evolution by Variation and Selection. }
\end{center}

In its initial formulation the theory was completely narrative.
Not a single mathematical equation was used. There have been constant
efforts by biologists and by others to clarify its meaning and
to make it more quantitative and hence more predictive
\cite{fisher1930, wright1932, li1955, li1977, kimura1983, 
maynardsmith1982, endler1986, svirezhev1990,hamilton1996, 
michod1999, barton2002, ewens2004, gellispie2004, waxman2005, ao2005, blows2007}.
Tremendous progress has been made during the past 100 years.
Two of the most important concepts emerged in Darwinian dynamics are
Fisher's fundamental theorem of natural selection \cite{fisher1930}, which
stresses the role of variation in evolution, 
and Wright's adaptive landscape \cite{wright1932}, which
describes the ultimate selection under a given condition as a potential function of
a landscape in a gigantic genetic space.
Nowadays the use of mathematics in this area is
comparable to that of any mathematically sophisticated field
of natural science. Darwinian dynamics is a {\it bona fide}
nonequilibrium stochastic dynamical theory, which governs the processes leading
to complex creatures such as {\it Methenobacterium} and {\it Homo sapiens} on Earth.

What would be a possible mathematical structure for Darwinian
dynamics? Though the scope of Darwinian dynamics is very broad and
its quantification appears formidable,
we will present here a precise and nevertheless general enough formulation.
Intuitively evolution is about successive processes:
Quantities at a later stage are related to
their values at its earlier stage under both predictable
(deterministic) and unpredictable (stochastic) constraints. For
example, the world population of humans in next 20 years will be
almost surely related to its current one. Hence, the genetic frequency,
the probability in the population, of a given form of gene
(allele) in the next generation is related to its present value.
Here sexual conducts and other reproduction behaviors are treated
as means to realize the variation and selection for evolution. We
may denote those genetic frequencies as ${\bf q}$ with $n$
components denoting all possible alleles. Thus ${\bf q}^{\tau} =
(q_1, q_2, ... , q_n)$ is a vector (Here $\tau$ denotes the
transpose). There are huge amount of human traits related to
genetics (or genes): height, skin color, size of eye ball, faster
runner, gene for liver cancer, gene for smartness, ... . The
number $n$ is then large: it could be as large, and likely larger, as the
number of genes in human genome, which is about 20000, if one
simply assigns one allele or a trait to one gene without any 
combinatory consideration. (This is certainly very crude. In
addition, we do not know the exact way to specify such
relationship yet). This number is far larger than the number of
chemical elements, which is about 100, and than the number of
elementary particles, about 30. With a suitable choice of time
scale equivalent to an averaging over many generations, the
incremental rate in such an evolutionary process may be
represented by a time derivative, ${\dot {\bf q}} = d {\bf q}/dt$.
The deterministic constraint at a given time may be represented by
a deterministic force ${\bf f}({\bf q}, t)$. For example, there is
a high confidence to predict the eye color of a child based on the
information from his/her parents, but the smartness of an
offspring is not so strongly correlated to that of the parents. The
random constraint, the unknown and/or irrelevant force,
is approximated by a
Gaussian-white noise term ${\bf \zeta}({\bf q}, t)$, with zero
mean, $\langle {\bf \zeta} \rangle = 0$ and the $n \times n$
variance matrix $D$: $ \langle {\bf \zeta}({\bf q}, t') {{\bf
\zeta}}^{\tau}({\bf q}, t) \rangle = 2 D({\bf q}, t) \theta \delta
(t-t') $. Here the factor 2 is a convention and $\theta$ is a
positive numerical constant reserved for the role of temperature
in physical sciences. $\delta(t)$ is the Dirac delta function.
With these notations we are ready to transform the word equation
into a precise mathematical equation, which now reads
\begin{eqnarray}
 {\dot {\bf q}} & = & {\bf f}({\bf q}, t) + {\bf \zeta}({\bf q}, t) \; .
   \nonumber
\end{eqnarray}

To physicists and chemists, this equation looks similar to an equation already known for 100 years 
\cite{vankampen1992, coffey2004}, proposed by Langevin 50 years
after Darwin and Wallace. There are some difficult issues in connection to the Langevin equation, such as the absence of detailed balance condition, to be discussed below.
To mathematicians as well as biologists, it is in the form of
standard formulation of stochastic differential equations \cite{doob1984, delmoral2004}. 
We will return to above equation as Eq.(1)
in next section. However, an immediate question arises:
while we may represent the variation in evolution by the variance
matrix $D$, where is Wright's adaptive landscape and the
corresponding potential function?

It is known that the deterministic force ${\bf f}({\bf q}, t)$ in general
cannot be related to a potential function in a straightforward way,
that is, ${\bf f}({\bf q}, t) \neq - D({\bf q}, t) \nabla \phi({\bf q}, t)$.
Here $\nabla = (\partial/\partial q_1,
\partial/\partial q_2, ... ,
\partial/\partial q_n)^{\tau} $ is the gradient operation in the
phase space formed by ${\bf q}$, and $ \phi({\bf q}, t)$ is a
scalar function. In fact, the existence of such a non-equality is
ubiquitous in nonequilibrium processes. It is the breakdown of detailed balance.
A nonequilibrium process typically has the following five qualitative characteristics:  \\
 1) dissipative,  $\nabla \cdot {\bf f}({\bf q}, t) \neq 0 $;  \\
 2) asymmetric, $\partial f_j({\bf q}, t)/ \partial q_i \neq
  \partial f_i({\bf q}, t)/ \partial q_j$ for at least one pair of indices of $i,j$; \\
 3) nonlinear, $ {\bf f}(\theta {\bf q}, t) \neq \theta \, {\bf f}({\bf q}, t)$;  \\
 4) stochastic with multiplicative noise, $D({\bf q}, t)$ depending on the state variable
  ${\bf q}$; and  \\
 5) possibly singular, that is, $\det( D({\bf q}, t) ) = 0$. \\
The asymmetry is the reason that ${\bf f}({\bf q}, t)$ cannot be
equal to $- D({\bf q}, t) \nabla \phi({\bf q}, t)$, which will
become explicit below. These are the main reasons that a
consistent formulation of nonequilibrium processes has been
difficult \cite{nicolis1977, haken1983, keizer1987, cross1993, langer1997}. 
Progress in recent biological studies has shown that a quantitative Wright adaptive
landscape is indeed embedded in the above stochastic differential
equation. It emerges in a manner completely consistent with its
use in the physical sciences \cite{ao2005, kwon2005, yin2006, zhu2006, ao2007},
which will be explored further in the present paper.

\subsection{Outstanding questions on statistical mechanics
and thermodynamics}

In the physical sciences there has been a sustained interest during
past several decades in nonequilibrium processes 
%
%Nicolis and Prigogine (1977), Haken (1983), Keizer (1987), van Kampen (1992),
%Cross and Hohenberg (1993), Risken (1996), Langer (1997), Gardiner (2004),
%Refregier (2004).
\cite{nicolis1977, haken1983, keizer1987, vankampen1992, risken1996, 
cross1993,langer1997,gardiner2004, refregier2004}.
Important goals are to bridge its connection to equilibrium
processes and to clarify the roles of entropy and the Second Law
of thermodynamics within deterministic and conserved dynamics
and to balance the descriptions between
the single particle trajectory and the ensemble distribution.
In this content we should also mention the enormous effort since
Boltzmann to understand the associated direction of time from
reversible dynamics \cite{zeh2007}.

There is also an active interest from the philosophical point of view
on the foundations of statistical mechanics and thermodynamics
\cite{sklar1993, emch2002}.
Relevant to the present paper, following three fundamental questions
have been explicitly formulated \cite{uffink2005}:
%\cite{uffink}
      \\
 "In what sense can thermodynamics said to be reduced to statistical
mechanics? \\
 "How can one derive equations that are not time-reversal
invariant from a time-reversal invariant dynamics? \\
 "How to provide a theoretical basis for the 'approach to
equilibrium' or irreversible processes?"

In the view of the absence of logical consistent answers to above questions
from conservative dynamics, Newtonian dynamics or quantum mechanics
\cite{gallavotti2004, cohen2005a},
it should be desirable to look into these problems from a completely
different perspective.

\subsection{What can we learn from Darwinian dynamics?}

Thanks to recent progress in experimental technologies,
particularly in nanotechnology, many previously inaccessible regimes in time
and space are now actively been exploring.
There have renewed interests in stochastic phenomena,
ranging from physics \cite{bustamante2005, deem2007, gallavotti2004, gaspard2007, gaveau2003,
hanggi2005,karevski2006, lebowitz1999, luchinsky1998, xing1998},
chemistry \cite{attard2007, qian2005a, reguera2005, sinitsyn2007},
material science \cite{beretta1984, ottinger2005, sasa2006}, 
biology \cite{ao2005, kaern2005, kussell2005, wang2006}, 
and to other fields \cite{delmoral2004}.
These work demonstrate the strong on-going exchange of ideas between
physical and biological sciences. In particular, quantitative experimental
and theoretical physical methodologies have been finding their way
into the study of cellular and molecular processes of life, which
has been very useful.
On the other hand, we think it is of at least equal value to consider
the opposite direction of the flow of ideas. Darwinian dynamics may
generate new insights in the physical sciences.

For example, Darwinian dynamics can address all three
fundamental questions in previous subsection in its own way.
For the first question, as long as
statistical mechanics is formulated according to the
Boltzmann-Gibbs distribution, it will be shown below that
Darwinian dynamics indeed implies this distribution, and that
the main structures of statistical mechanics and thermodynamics are
equivalent. For the second question, it is found that
thermodynamics is based on the energy conservation (the First Law)
and on the Carnot cycle. It deals with quantities at
equilibrium or steady state. There is no role for dynamics. Thus,
there is no requirement for the direction of time: Both
conservative and nonconservative dynamics can be consistent with
it. The explicit independence of the Carnot cycle and the
First Law on dynamical properties in Darwinian dynamics will
become clear below. Hence, there is no conflict between
thermodynamics and the time-reversal dynamics dominated in
physics. For the third and last question, Darwinian dynamics comes
with an adaptive behavior \cite{ao2005, blows2007, darwin1958, darwin1858, li1955, endler1986, ewens2004, fisher1930, gellispie2004, maynardsmith1982, michod1999, waxman2005, wright1932}
and with an intrinsically built-in direction of time. This is due
to the explicitly stochastic or probabilistic nature of Darwinian
dynamics. Such a behavior was summarized as the fundamental
theorem of natural selection \cite{fisher1930} and extended further
as the F-theorem \cite{ao2005}. Thus, Darwinian dynamics provides a
general framework to address the question of "approaching to
equilibrium".

\subsection{Organization of the paper}

The rest of the paper is organized as follows. In section II
Darwinian dynamics will be summarized. In section III it will be
shown that statistical mechanics and the canonical ensemble follow
naturally from Darwinian dynamics. In section IV the connection to
thermodynamics is explored. There it will be shown that the Zeroth
Law, the First Law, and the Second Law follow directly and naturally from
Darwinian dynamics, but not the Third Law. In section V two types
of simple, but seemly profound, dynamical equalities discovered
recently are discussed. One is based on the Feynman-Kac formula
and one is a generalization of Einstein relation. In section VI
the range of present ideas is put into perspective.

Two disclaimers should be made at the beginning of our
discussion. First, we will mainly be concerned with the
theoretical structures, not with specific details. Specifically,
we will focus on which structural elements should be presented in
various equations, where consensus can be reached, not so much as
what would be details forms of each elements. In fact, many of
detailed forms are still unknown and are active research topics.
Second, rigorous mathematical proofs will be not provided, though
care has been taken to make presentations as clear and consistent as
possible.
%With the solid physical and
%biological foundations a rigorous mathematical formulation is
%nevertheless possible.

%
%universal (thermodynamics) vs open (randomness)
%
% Boltzmann: realization, large but single system,
%    approximate procedure
% $N \rightarrow \infty$    vs
% Gibss: idealization, probability,
%    explanatory principle.
%

\section{ Darwinian Dynamics, Adaptive Landscape, and F-Theorem }

This section summarizes recent results on Darwinian
dynamics, in more detail than provided in the Introduction.

\subsection{ Stochastic differential equation: the particle and trajectory view }

In the context of modern genetics Darwin's theory of evolution
may be summarized verbally  as "the evolution is a result of
genetic variation and its ordering through elimination and
selection". Both randomness and selection are equally important in
this dynamical process, as encoded into Fisher's fundamental
theorem of natural selection \cite{fisher1930} and Wright's adaptive
landscape \cite{wright1932}. With an appropriate time scale,
Darwinian dynamics may be represented by the following stochastic
differential equation \cite{ao2005, li1977, svirezhev1990}
\begin{equation}  \label{standard}
  \dot{{\bf q}} = {\bf f}({\bf q}) + N_I({\bf q}) {\bf \xi}(t) \; ,
\end{equation}
where ${\bf f}$ and ${\bf q}$ are $n$-dimensional vectors and
${\bf f}$ a nonlinear function of ${\bf q}$. The genetic frequency
of $i$-th allele is represented by $q_i$. Nevertheless, in the
present paper it will be treated as a real function of
time $t$. Depending on the situation under consideration,
the quantity ${\bf q}$ could, alternatively, be the populations of
$n$ species in ecology, or, the $n$ coordinates in physical
sciences. All quantities in this paper are dimensionless unless
explicitly specified. They are assumed to be measured in their own
proper units. The collection of all ${\bf q}$ forms a real
$n$-dimensional phase space. The noise ${\bf \xi}$ is explicitly
separated from the state variable to emphasize its independence,
with $l$ components. It is a standard Gaussian white noise function
with $ \langle  \xi_i \rangle_{\xi} = 0 \; ,
$
and
\begin{equation}
 \langle \xi_i(t) \xi_j (t')\rangle_{\xi}
  = 2 \theta \; \delta_{ij} \delta (t-t') \; ,
\end{equation}
and $i,j=1, 2, ..., l$. Here $ \langle ... \rangle_{\xi} $ denotes
the average over the noise variable $\{{\bf \xi} (t) \}$, to be
distinguished from the average over the distribution in phase
space below. The positive numerical constant $\theta$ describes
the strength of noise. The variation is described by the noise
term in Eq.(\ref{standard}) and the elimination and selection
effect is represented by the force ${\bf f}$.

A further description of the noise term in Eq.(\ref{standard}) is
through the $n\times n$ diffusion matrix $D({\bf q})$, which is
defined by the following matrix equation
\begin{equation}
  N_I({\bf q}) N_I^\tau ({\bf q}) = D({\bf q}) \; ,
\end{equation}
where $N_I$ is an $n\times l$ matrix,  $N_I^\tau$ is its the
transpose, which describes how the system is coupled to the noisy
source. This is the first type of the F-theorem \cite{ao2005},
a generalization of Fisher's fundamental theorem of natural
selection \cite{fisher1930} in population genetics. 
According to Eq.(2) the $n\times n$
diffusion matrix $D$ is both symmetric and nonnegative. For the
dynamics of the state vector ${\bf q}$, all that is needed from the noise
term in Eq.(1) are the diffusion matrix $D$ and the positive
numerical parameter $\theta$. Hence, it is not even necessary to
require the dimension of the stochastic vector $\xi$ be the same
as that of the state vector ${\bf q}$. This implies that in
general $l \neq n$.

We emphasize here that an extensive class of nonequilibrium processes can
indeed be described by such a stochastic differential equation
\cite{nicolis1977, haken1983, keizer1987, vankampen1992, risken1996, gardiner2004, coffey2004}.
The current research efforts on such stochastic and
probability description are ranging from physics \cite{bustamante2005, hanggi2005},
chemistry \cite{reguera2005, qian2005a},
material science \cite{ottinger2005, sasa2006},
biology \cite{ao2005, kaern2005, kussell2005, wang2006},
and other fields \cite{delmoral2004}.

Darwinian dynamics was conceived graphically by Wright \cite{wright1932}
as the motion of the system in an adaptive landscape in genetic
space (for an illustration, see Fig. 1).
Since then such a landscape has been known as the fitness
landscape in some part of literature \cite{kauffman1993, michod1999, waxman2005}. 
However, there exists a considerable amount of confusion about the definition of fitness
\cite{endler1986, michod1999, ao2005}.
In the present paper a more neutral term, the (Wright
evolutionary) potential function, will be used to denote this
landscape. The adaptive landscape connecting both the individual
dynamics and its final destination is intuitively appealing.
Nevertheless, it has been difficult to prove its existence in a
general setting. The major difficulty lies in the fact that typically
the detailed balance condition does not hold in Darwinian
dynamics, that is, $D^{-1}({\bf q}) {\bf f}({\bf q})$ cannot be
written as a gradient of scalar function \cite{nicolis1977,
vankampen1992, cross1993, risken1996, gardiner2004}, 
already mentioned in the Introduction.

%\begin{turnpage}
%\begin{figure}
%% \epsfig{file=fig1.eps,width=3.65in}
%% \epsfysize=4in
%% \epsfbox{fig1.eps}
% \includegraphics{fig1.eps}
%% \caption{ \label{fig1}
%%  Adaptive landscape with in potential contour representation.
%%  $+$: local basin; $-$: local peak;
%%  $\times $: pass (saddle point).}
%\end{figure}
%\end{turnpage}

\begin{figure}
% \epsfig{file=fig1.eps,width=3.65in}
% \epsfysize=4in
% \epsfbox{fig1.eps}
% \includegraphics{fig1.eps}
For the figure, please check the published version: Emerging of Stochastic Dynamical Equalities and Steady State Thermodynamics from  Darwinian Dynamics, P. Ao, Communications in Theoretical Physics {\bf 49} (2008) 1073-1090. 
 \caption{ \label{fig1}
  Adaptive landscape with in potential contour representation.
  $+$: local basin; $-$: local peak;
  $\times $: pass (saddle point).}
\end{figure}

During our study of the robustness of the genetic switch in a
living organism \cite{zhu2004, zhu2007}
a constructive method was discovered to overcome those difficulties:
Eq.(1) can be transformed into the following stochastic
differential equation,
\begin{equation} \label{normal}
  [ R({\bf q}) + T({\bf q})] \dot{{\bf q}} = - \nabla
   \phi({\bf q}; \lambda) + N_{II}({\bf q})\xi(t) \; ,
\end{equation}
where the noise $\xi$ is from the same source as that in Eq.(1).
The parameter $\lambda$ denotes the influence of non-dynamical and
external quantities.  It should be pointed out that the potential
function $\phi$ may also implicitly depend on $\theta$. The
friction matrix $R({\bf q})$ is defined through the following
matrix equation
\begin{equation}
  N_{II}({\bf q}) N_{II}^\tau ({\bf q})= R({\bf q}) \; ,
\end{equation}
which guarantees that $ R $ is both symmetric and nonnegative.
This is the second type of the F-theorem \cite{ao2005}.
The F-theorem emphasizes the connection between adaption and
variation and is essentially a reformulation of
fluctuation-dissipation theorem in physics \cite{degroot1984, kubo1992, zwanzig2001}
and of Fisher's fundamental theorem of natural selection. The connection
between Fisher's fundamental theorem of natural selection and the
fluctuation-dissipation theorem was also noticed recently by others \cite{sato2003}. 
It should be emphasized here that the F-theorem is not confined to the neighborhood of an equilibrium or
steady state. It is valid in nonlinear cases without detailed
balance: There is no reference to potential function in the
definition of friction matrix $R$ and the anti-symmetric matrix
$T$ is in general nonzero.
For simplicity we will assume $\det( R ) \neq 0$ in the rest of
the paper. Hence $\det( R + T ) \neq 0$ \cite{kwon2005}.
The breakdown of detailed balance condition or the time reversal
symmetry is now represented by the finiteness of the transverse
matrix, $T \neq 0$. The usefulness of the formulation of
Eq.(\ref{normal}) has already been demonstrated in the successful solution
of an outstanding stability puzzle in gene regulatory dynamics \cite{zhu2004, zhu2007}
and in a consistent formulation of Darwinian dynamics \cite{ao2005}.
Evidently, the Wright adaptive landscape and the F-theorem are the
realization of chance and necessity in evolution \cite{monod1971}.

The $n\times n$ symmetric, non-negative "friction matrix" $ R $ and
the "transverse matrix" $T$ are directly related to the diffusion matrix $D$:
$$
%\begin{equation}
 R({\bf q}) + T({\bf q})
  = {1 \over { D({\bf q}) + Q({\bf q})} } \; .
%\end{equation}
$$
Here $Q$ is an antisymmetric matrix determined by both the
diffusion matrix $D({\bf q})$ and the deterministic force ${\bf
f}({\bf q})$ \cite{ao2004, kwon2005}.
One of more suggestive forms of above equation is
\begin{equation} \label{einstein}
 [ R({\bf q}) + T({\bf q}) ] D({\bf q}) [ R({\bf q}) - T({\bf q}) ]
   = R({\bf q}) \; .
\end{equation}
This symmetric matrix equation implies $n(n+1)/2$ single equations
from each of its elements. We need another $n(n-1)/2$ equations
in order to completely determine the matrices $R$ and $T$,
which will come from the conditions for the potential function.

The Wright evolutionary potential function $\phi({\bf q})$ is connected
to the deterministic force ${\bf f}({\bf q})$ by
$$
%\begin{equation}
 - \nabla \phi({\bf q}; \lambda)
   =  [ R({\bf q}) + T({\bf q}) ] {\bf f}({\bf q}) \; .
%\end{equation}
$$
Or its equivalent form,
\begin{equation} \label{p-condition}
 \nabla \times [ [R({\bf q}) + T({\bf q}) ] {\bf f}({\bf q}) ] = 0 \; .
\end{equation}
Here the operation $\nabla \times$ on an arbitrary $n$-dimensional
vector ${\bf v}$ is a matrix generalization of the curl operation
for lower dimensions ($n=2,3$): $
 (\nabla \times {\bf v} )_{i,j} = \partial  v_j /\partial q_i
- \partial  v_i/\partial q_j \; . $ Above matrix equation is hence
antisymmetric and gives the needed $n(n-1)/2$ single equations
from each of its elements. From Eq.(\ref{einstein}) and
(\ref{p-condition}) the friction matrix, $R$, the transverse
matrix, $T$, and the potential function, $\phi$, can be
constructed from the diffusion matrix $D$ and the deterministic
force ${\bf f}$. The boundary condition in solving
Eq.(\ref{p-condition}) is implied by the requirement that the
fixed points of ${\bf f}$ should coincide with the extremals of
the potential function $\phi$. The local construction, the
construction near any fixed point, was demonstrated in detail in Ref.
\cite{kwon2005}, where the connection to the
fluctuation-dissipation theorem in physical sciences was
explicitly demonstrated.
For the global construction valid in the whole phase space
an iterative method was outlined in Ref. \cite{ao2004}.
Some of its mathematical and properties, such as the speed of convergence,
are not generally known at this moment.

For the case where the stochastic drive may be ignored, that is, $\theta
= 0 $, the relationship between Eq.(1) and (\ref{normal}) remains
unchanged, but Eq.(\ref{normal}) becomes deterministic
\begin{equation} \label{adaptive}
  [ R({\bf q}) + T({\bf q})] \dot{{\bf q}} = - \nabla
   \phi({\bf q}; \lambda)  \; .
\end{equation}
The nonlinear dynamics typical in evolutionary processes is usual
explored in the framework of game theory \cite{hofbauer1998, maynardsmith1982}. 
The typical mathematical equation is 
of the form in Eq.(1) without noise: $\dot{{\bf q}} = {\bf f}({\bf
q}; \lambda)$. The universal construction of Lyapunov function in
the game theory had been an unsolved problem before the present
formulation. On the other hand, above equation indicates, with the
non-negativeness of the friction matrix,
\begin{eqnarray}
 \frac{d }{d t} \phi({\bf q}; \lambda)
   & = & \dot{{\bf q}} \cdot  \nabla \phi({\bf q}; \lambda)
   \nonumber \\
   & = & - \dot{{\bf q}}^{\tau}  [ R({\bf q}) + T({\bf q}) ]
    \dot{{\bf q}} \nonumber \\
   & = & - \dot{{\bf q}}^{\tau}  R({\bf q}) \dot{{\bf q}}
   \nonumber \\
   & \leq &  0 \; .
\end{eqnarray}
It is clear then that the Wright evolutionary potential
function $\phi({\bf q}; \lambda)$ is a Lyapunov function.
The deterministic dynamics makes it non-increasing, and it
approaches the nearby potential minimum to achieve the maximum
probability. This is precisely what was conceived by Wright.
Adaptive dynamics is actively been exploring in biology \cite{blows2007,
waxman2005}.

The idea of potential function landscape has a long history in biological sciences.
Such an idea was first proposed in population genetics \cite{wright1932}.
It was proposed again in developmental biology as a developmental landscape
\cite{waddington1940}, and again in the description of a genetic
switch in molecular biology \cite{delbruck1949}. Similar landscape
idea was proposed in the ecological evolution embedded in the
concept of ascendency \cite{ulanowicz1980}. The landscape has been
used not only to model neural computation \cite{hopfield1984, hertz1991}, 
also to understand the protein folding dynamics \cite{wolynes2005}. 
The problem of absence of detailed balance, however,  had
been previously regarded as an obstacle. For example, it was noted
that genuine nonequilibrium and asymmetric dynamics such as limit
cycle might make the construction of the Hopfield potential
function in neural computation impossible. Recently, it was shown
this can be overcome by the present formulation \cite{zhu2006}.

Conservative Newtonian dynamics may be regarded as another
limit of the above formulation: zero friction limit, where $ R=0 $ in addition to $\theta = 0$.
Hence, from Eq.(\ref{adaptive}), Newtonian dynamics may be expressed as,
\begin{equation} \label{newton}
  T({\bf q}) \; \dot{{\bf q}} = - \nabla  \phi({\bf q}; \lambda)  \; .
\end{equation}
Here the value of potential function is evidently conserved during the
dynamics since  $\dot{{\bf q}} \cdot  \nabla \phi({\bf q}; \lambda) = 0 $.
The system moves along equal potential contours in the adaptive landscape.
This conservative behavior suggests that the rate of approaching to
equilibrium is associated with the friction matrix $R$,
not with the diffusion matrix $D$.
There are situations where the diffusion matrix is
finite but the friction matrix is zero, and thus the dynamics is
conservative \cite{zhu2006}.

\subsection{Fokker-Planck equation: the ensemble and distribution view }

It was reasoned heuristically \cite{ao2004}
that the steady state distribution $\rho({\bf q})$ in the state
space, if exists, is
\begin{equation} \label{bg}
 \rho({\bf q},t=\infty) \propto
  e^{ - \beta {\phi({\bf q}; \lambda ) } } \; .
\end{equation}
Here $\beta = 1/\theta$. It takes the form of Boltzmann-Gibbs
distribution function. Therefore, the potential function $\phi$
acquires both the dynamical meaning through Eq.(\ref{normal}) and
the steady state meaning through Eq.(\ref{bg}).

It was further demonstrated that such a heuristic argument can
be translated into an explicit algebraic procedure such that there is an
explicit Fokker-Planck equation whose steady state solution is
indeed given by Eq.(\ref{bg}) \cite{yin2006}.
Starting with the generalized Klein-Kramers equation, taking
the limiting procedure of the zero mass limit, the desired
Fokker-Planck equation corresponding to Eq.(\ref{normal}) is
\begin{equation} \label{fp-eq}
 {\partial \rho({\bf q},t) \over \partial t}
  = \nabla^{\tau} [D({\bf q}) + Q({\bf q}) ] [\theta \nabla
   + \nabla \phi({\bf q}; \lambda)] \rho({\bf q},t) \; .
\end{equation}
This equation is equivalent to a statement of conservation of probability.
It can be rewritten as the probability continuity equation:
\begin{equation} \label{continuity}
 {\partial \rho({\bf q},t) \over \partial t}
  + \nabla \cdot {\bf j}({\bf q}, t) = 0  \, ,
\end{equation}
with the probability current density ${\bf j}$ defined as
\begin{equation} \label{current}
 {\bf j}({\bf q}, t) \equiv - [D({\bf q}) + Q({\bf q}) ]
   [\theta \nabla + \nabla \phi({\bf q}; \lambda)]
   \rho({\bf q},t) \; .
\end{equation}
The reduction of dynamical variables has often been done in the
well-known Smoluchowski limit. In the above derivation we take the
mass to be zero, keeping other parameters, including the friction
and transverse matrices, to be finite.
In the Smoluchowski limit, however, the friction matrix is taken to
be infinite, keeping all other parameters finite. Those two limits
are in general not exchangeable.

The steady state configuration solution of Eq.(\ref{fp-eq}) is
indeed given by Eq.(\ref{bg}). It is interesting to point out
that the steady state distribution function, Eq.(\ref{bg}), is
independent of both the friction matrix $ R $ and the transverse
matrix $ T $. Furthermore, we emphasize that no detailed balance
condition is assumed in reaching this result. In addition, both
the additive and multiplicative noises are treated here on an equal
footing.

Finally, it can be verified that above construction leading to
Eq.(\ref{fp-eq}) is valid, and remains unchanged, when there is an
explicit time dependence in $R$, $T$, and/or $\phi$. There may not
exist a steady state distribution, for example, if the Wright
evolutionary potential function $\phi$ depends on time.

\section{ Nonequilibrium Statistical Mechanics }

\subsection{Central relation in statistical mechanics}

If we treat the parameter $\theta (= 1/\beta ) $ as temperature,
the steady state distribution function in phase space
is indeed the familiar Boltzmann-Gibbs distribution,
Eq.(\ref{bg}). The partition function, or the normalization
constant, is then
\begin{equation} \label{p-function}
 {\cal Z}_{\theta}(\lambda) \equiv \int d^n {\bf q} \,
  e^{ - \beta {\phi({\bf q}; \lambda) } } \; .
\end{equation}
The integral $\int d^n {\bf q}$ denotes the summation over whole
phase space. The normalized steady state distribution is
\begin{equation}
 \rho_{\theta}({\bf q})
  \equiv  \frac{e^{ - \beta {\phi({\bf q}; \lambda) } } }
   { {\cal Z}_{\theta} } \; .
\end{equation}
For a given observable quantity $O({\bf q})$, its average or
expectation value is
\begin{eqnarray} \label{summit}
 \langle O \rangle_{\bf q}
  & \equiv &  \int d^n {\bf q} \, O({\bf q}) \,
       \rho_{\theta}({\bf q}) \nonumber \\
  & = &  \frac{1}{{\cal Z}_\theta} \int d^n {\bf q} \, O({\bf q}) \,
  e^{ - \beta {\phi({q}; \lambda) } } \: .
\end{eqnarray}
The subscript $q$ denoted that the average is over phase space,
not over the noise in Eq.(\ref{standard}) or (\ref{normal}).
Eq.(\ref{summit}) is the main fortress of statistical mechanics.
Working in statistical mechanics then may be classified into two
types: conquering the fortress from outside, that is, formulating
as instance of Eq.(\ref{summit}); and conquering more territory
from the fortress, that is, applying Eq.(\ref{summit}).

There has been tremendous amount of efforts to derive the
canonical distribution of Eq.(16) from conservative dynamics. One of
the best results is the typicality of such distribution for large
systems already attempted by Boltzmann \cite{goldstein2006}. 
On the other hand, all experiments have shown a universal validity of Eq.(16) for both large and small systems, hence more than typicality. We note that
Darwinian dynamics is consistent with such empirical observations.

There is a difference in the use of potential function $\phi$ in Eq.(12) and  Eq.(15): One is in the form of "force"--gradient with respect to the coordinate $\bf q$, and another is its integration which may carry an arbitrary function of the parameter $\lambda$: 
$\phi({\bf q}; \lambda) = \phi_0({\bf q}; \lambda) + \phi_1(\lambda)$. For a static parameter, this would not be of any problem: It simply reflects the fact that only the difference of the potential function with respect to a given reference is meaningful. Nevertheless, if we are going to compare the potential function at two different parameter values,  $\phi({\bf q}; \lambda_1)$ and 
$\phi({\bf q}, \lambda_2)$, connected by a dynamical processes controlled by $\lambda(t)$, 
such an arbitrary function $\phi_1(\lambda)$ has to be fixed up to a constant. Otherwise, the free energy to be discussed in Sec. IV and V would be arbitrary. For a conservative dynamics described by Eq.(10), this function may be determined by a procedure named the minimum gauge condition: Assume an adiabatic (slow) process connecting two states specified by $\lambda_1$ and $\lambda_2$, and let $\delta {\bf f} ({\bf q}; \lambda(t)) = - \nabla [ \phi({\bf q}; \lambda(t)) - \phi({\bf q}; \lambda = 0) ]$, the force directly controlled by the parameter $\lambda$, the minimum gauge condition to determine $\phi_1(\lambda)$ may be expressed as
\[
  \phi_1(\lambda_2) - \phi_1(\lambda_1) = \left. W \right|_{t_2-t_1 \rightarrow \infty } \; ,
\]    
and $ W = \int_{\lambda(t=t_1) = \lambda_1, \lambda(t=t_2) = \lambda_2 } 
d{\bf q} \cdot \delta {\bf f} ({\bf q};\lambda(t))$, that is, the work done in the adiabactic process by the external force related to the parameter is equal to the change in potential function, a known relation in classical mechanics.

\subsection{Stochastic processes and the canonical ensemble}

A fundamental question raised by our formulation is this: for a
given Fokker-Planck equation, can the corresponding stochastic
differential equation in the form of Eq.(\ref{normal}) be
recovered (the inverse problem)? That is, is there a one-one correpodence between 
the local and global dynamics connected by a potential function? 
The answer is yes, and the
procedure for carrying it out is implicitly contained in Eq.(\ref{fp-eq}),
which will be demonstrated below.

A generic form for the Fokker-Planck equation is expressed as
follows:
\begin{equation} \label{fp-generic}
  {\partial \rho({\bf q},t) \over \partial t}
  = \nabla^{\tau} [\theta  \overline{D}({\bf q}) \nabla
    - \overline{{\bf f}}({\bf q})] \rho({\bf q},t) \; .
\end{equation}
Here ${\overline{D}}({\bf q})$ is the diffusion matrix and
$\overline{{\bf f}}({\bf q})$ the drift force. The main motivation
for such a form is simple: In the case detailed balance condition
is satisfied, {\it i.e.}, $Q({\bf q}) = 0$ (and $T({\bf q}) = 0$),
the potential function $\overline{\phi}$ can be directly inferred
from above equation: $\nabla \overline{\phi} = \overline{D}^{-1}
\; \overline{\bf f}$. This makes the diffusion effec very
prominent. Any other form of the Fokker-Planck equations can be
easily transformed into the above form. This generic form of the
Fokker-Planck equation is less amenable to additional
complications such as the noise induced phase transitions caused
by the ${\bf q}$-dependent diffusion constant.

A potential function $\overline{\phi}({\bf q})$ can always be
defined from the steady state distribution. There is an extensive
mathematical literature addressing this problem \cite{doob1984}.
After this is done, though it can be a difficult problem,
the procedure to relate the Fokker-Planck equation to Eq.(\ref{fp-eq})
is straightforward. Eq.(\ref{fp-eq}) can be rewritten as
\begin{equation}  \label{fp-eq2}
 {\partial \rho({\bf q},t) \over \partial t}
  = \nabla^{\tau} [\theta D({\bf q})\nabla
   + \theta (\nabla^{\tau} Q({\bf q}))
   +  [D({\bf q}) + Q({\bf q})]\nabla \phi({\bf q})]
      \rho({\bf q},t) \; .
\end{equation}
The antisymmetric property of the matrix $Q({\bf q})$ has been
used in reaching Eq.({\ref{fp-eq2}).  Thus, comparing
Eq.(\ref{fp-generic}) and (\ref{fp-eq2}), we have
\begin{eqnarray}
  D({\bf q})
   & = & \overline{D}({\bf q}) \; , \\
  \phi({\bf q})
   & = & \overline{\phi}({\bf q}) \; , \\
  {\bf f}({{\bf q}})
   & = & \overline{{\bf f}}({\bf q})
         - \theta \nabla^{\tau} Q({\bf q}) \; ,  \label{force}
\end{eqnarray}
where we have used the relation
$$
 - [D({\bf q}) + Q({\bf q})] \nabla \phi({\bf q})
   = {\bf f}({\bf q}) \; .
$$
The explicit equation for the anti-symmetric matrix $Q$ is
\begin{equation} \label{Q-eq}
  - \theta \nabla^{\tau} Q({\bf q})
   - [ D({\bf q}) + Q({\bf q}) ]  \nabla \phi({\bf q}; \lambda)
    =  \overline{{\bf f}}({\bf q}) \; ,
\end{equation}
which is a first order, linear, inhomogeneous, partial differential
equation. The solution for $Q$ can be formally written as
\begin{equation}
  Q({\bf q}) = - {1 \over{\theta}} \int^{\bf q} d{\bf q}'
   [ \overline{\bf f}({\bf q}')
    + D({\bf q}')\nabla ' \phi({\bf q}'; \lambda) ]
    e^{ \beta ({\phi({\bf q}; \lambda)  - \phi({\bf q}'; \lambda)} ) }
     + Q_0({\bf q}) e^{  \beta \phi({\bf q}; \lambda ) }  \; .
\end{equation}
Here $Q_0({\bf q})$ is a solution of the homogenous equation
$\theta \nabla^{\tau} Q({\bf q}) = 0 $, and the two parallel
vectors in the integrand, $d{\bf q}' \; \overline{\bf
f}({\bf q})$, defines a matrix. This completes our answer to the
inversion problem.

It is interesting to note that the shift between the zero's of the
potential gradient and the drift is given from
Eq.(\ref{force}) as,
\begin{equation} \label{shift}
 \Delta \overline{\bf f}
  = - \theta \nabla^{\tau} Q({\bf q}) \; .
\end{equation}
The extremals of the steady state distribution are not
necessarily determined by the zero's of drift.
This shift can occur even when $D=constant$.
The indication for such a shift appeared extensive in numerical studies 
\cite{lindner2004}. It was also noted analytically \cite{zhou1980}.

Thus, the zero-mass limit approach to the stochastic differential
equation is consistent in itself. The meaning of the potential,
$\phi$, is explicitly manifested in both the local trajectory, according
to Eq.(\ref{normal}), and the ensemble distribution, according to
Eq.(\ref{fp-eq}). In particular, no detailed balance condition is
assumed. There is no need to differentiate between the additive
and multiplicative noise. This zero mass limit procedure which
leads to Eq.(\ref{fp-eq}) from Eq.(\ref{normal}) may be regarded
as another prescription for stochastic integration, in
addition to those of Ito \cite{risken1996, gardiner2004}, of
Stratonovich-Fisk \cite{fisk1965, vankampen1992}, and of
Hanggi-Klimontovich \cite{hanggi1978, klimontovich1994}, and of others
\cite{wong1965}. They have been discussed from a unified
mathematical perspective in terms of an initial-point, middle-point, and end-point
discretization rules \cite{wong1965}.
All those previous methods of treating the stochastic differential equation are mathematically consistent in themselves and are related to each other.
The connection of the present method to those previous methods is suggested by
Eqs.(\ref{fp-generic}) and (\ref{fp-eq}) (or Eq.(\ref{fp-eq2}) ).
For example, Eq.(18) is just what can be obtained from the Hanggi-Klimontovich
type treatment, which has been noticed by others as well \cite{graham1971, eyink1996}. 
It is interesting to note that Ito's method puts an emphasis on
the martingale property of stochastic processes, which may be
viewed as a prescription from mathematics. The Stratonovich-Fisk
method stresses the differentiability such that the usual
differential chain-rule can be formally applied, which may be
viewed as the prescription from engineering. The
Hanggi-Klimontovich type stresses the generalized detailed
balance, important in physics. The present approach emphasizes the
role played by the potential function in both trajectory and
ensemble descriptions, as well as the existence of a generalized Einstein
relation (see below) when the detailed balance is absent. It may
be regarded as the prescription from natural sciences. All those
stochastic integration methods point to the need for an explicit
partition between the stochastic and deterministic forces, the
hallmark of hierarchical structure in dynamics. This feature
corresponds precisely to the hierarchical law in the evolutionary
dynamics of biology \cite{ao2005}.

Two more remarks are in order here. 
First, by construction the present method preserves the fixed points: The fixed points of $\bf f$ are also those of $\nabla \phi$. The introducing of the stochastic force would not shift the fixed points. This is very useful in that, the results of powerful bifurcation analysis of deterministic dynamics can be directly carried over to the stochastic situation. Second, there is a one to one correspondence between Eq.(4) and the dynamical equation in dissipative quantum phenomena \cite{leggett1992}. Because the latter has been discussed in context beyond white noise, this connection suggests an immediate generalization of Eq.(4) to colored noise situations.

We may conclude that a stochastic process
% regardless of Ito,
%Stratonovich-Fisk, Hanggi-Klimontovich, or the present method, or
%others,
%
leads to the canonical ensemble with a temperature and a
Boltzmann-Gibbs type distribution function, independent of how it is treated.
Other related stochastic ensembles, such as the grand canonical ensemble, may be
introduced in the same way by including additional constraints.

\subsection{Discrete stochastic dynamics}

There is another kind of modelling predominant in population
genetics and other fields which is discrete in phase space and/or
time. The existence of potential function in such stochastic
dynamical systems has been convincingly argued \cite{hill2005, qian1979, zia2007}. 
Here we will not discuss it in any detail, and simply quote a few relevant results.
The reasons to do so are: \\
1) It is known mathematically that any discrete model can be
represented exactly by a continuous one according to the embedding
theorem \cite{sauer1991, garay2001, lee2003},
though sometimes such a process may
turn a finite dimension problem into an infinite dimension one; \\
2) By a coarse graining, averaging process, the discrete dynamics in population
genetics can often be simplified to continuous ones such as
diffusion equations or Fokker-Planck equations \cite{li1977,
oppenheim1977, vankampen1992, gaveau1996, gardiner2004}.
It is generally acknowledged in population genetics and in other
fields that the diffusion approximation is often a good approximation.

For the steady state distribution, all one needs to know is the potential function $\phi$. 
The temperature can be set to be unity: $\theta = 1$. 
Hence, despite possible additional mathematical issues, the discrete or continuous representation
does not seem to be a physically or biologically important issue.

\section{ Steady State Thermodynamics }

Given the Boltzmann-Gibbs distribution, the partition function can
be evaluated according to Eq.(\ref{p-function}). Hence, in steady
state, all observable quantities are known in principle according
to Eq.(\ref{summit}). One may wonder then what can we learn about
a system from thermodynamics. First, there is a practical value.
In many cases the calculation of the partition function is hard,
if possible. It would be desirable if there are alternatives.
Thermodynamics gives us a set of useful relations between
observable quantities based on general properties of the system,
such as symmetries. Precise information on one observable can be
inferred from the information on other observables. Second, there
is a theoretical value. Thermodynamics has a scope far more
general than many other fields in physics. It is the only field in
classical physics whose foundation and structure not only have
survived quantum mechanics and relativity, but become stronger.
Furthermore, thermodynamics exhibits a sense of formal beauty,
elegance, and simplicity, which is exceedingly satisfying
aesthetically. Its influence is far beyond physical sciences, because
it is also based on probability and statistics.

There are numerous excellent books deriving thermodynamics from
statistical mechanics. A thorough treatment can be found, for
example, in Ref. \cite{callen1985}.
A more reader-friendly treatment can be found in Ref. \cite{ma1985} or
\cite{reichl1998}.
Concise and elementary treatments from thermodynamics point of
view have been presented in Ref. \cite{pippard1964}
and in Ref. \cite{reiss1996}.
A modern discussion of the approaching to the steady state was
presented in Ref. \cite{mackey1992}.
Ref. \cite{oono1998}
gave a comprehensive review from the point of view steady state thermodynamics,
but "temperature" was de-emphasized.
The present demonstration overlaps with it at various places. 
Nevertheless, there is one main difference: The role of "temperature" is explicitly discussed here.
Ref. \cite{sekimoto1998} gave a detailed discussion of the connection between
thermodynamics and Langevin dynamics with an emphasis on detail balance and
on the stochastic integration of Stratonovich.
The above demonstration already indicates that there is no need to confine to
Stratonovich approach.

The principal objective in this section is to show that Darwinian
dynamics indeed implies the main structures of thermodynamics,
even though at a first glance it seems to have no connection, because
Darwinian dynamics is at the extreme end of nonequilibrium processes.
In the light of those superb expositions mentioned above, the present discussion
may appear incomplete as well as arbitrary. For a systematic
discussion on thermodynamics the reader is sincerely encouraged to
consult those books and/or any of her/his favorites.
Nevertheless, we wish to show that a logically consistent dynamical understanding of thermodynamics
can be obtained.
Specifically, it is explicitly demonstrated that absence of detailed balance condition does not prevent us to obtain thermodynamics.

\subsection{Zeroth Law: existence of absolute ''temperature''}

From Darwinian dynamics, the steady state distribution is given by
a Boltzmann-Gibbs type distribution, Eq.(\ref{bg}), determined by
the Wright evolutionary potential function $\phi$ of the system
and a positive parameter $\theta$ of the noise strength. Hence,
the analogy of the Zeroth Law of thermodynamics is implied by
Darwinian dynamics: There exists a temperature-like quantity,
represented by the positive parameter $\theta$. This "temperature"
$\theta$ is "absolute" in that it does not depend on the system's
material details.
It is evident that the existence of the "temperature" is a direct
consequence of stochasticity in Darwinian dynamics, as exemplified in Eqs.(1-6).

%
%???? proportional coefficient is the Boltzmann constant
%The third univeral constant in physics (other two $\hbar$, $c$)
%
%The gravitational constant $G$ is different.
%

\subsection{First Law: conservation of ''energy''}

\subsubsection{Fundamental relation and the differential forms}

From the partition function ${\cal Z}_{\theta}$, we may define a
quantity
\begin{equation} \label{f-energy}
 F_{\theta} \equiv - \theta \ln {\cal Z}_{\theta} \; .
\end{equation}
We may also define the average Wright evolutionary potential
function,
\begin{equation} \label{i-energy}
 U_{\theta} \equiv \int d^n {\bf q} \; \phi({\bf q}; \lambda) \;
   \rho_{\theta}({\bf q}) \; .
\end{equation}
From the distribution function we may further define a positive
quantity
\begin{equation} \label{entropy}
 S_{\theta} \equiv - \int d^n {\bf q} \;
  \rho_{\theta}({\bf q}) \ln \rho_{\theta}({\bf q})  \; .
\end{equation}
It is then straightforward to verify that
\begin{equation}\label{first}
 F_{\theta}= U_{\theta} - \theta \; S_{\theta} \; ,
\end{equation}
precisely the fundamental relation in thermodynamics satisfied by
free energy, $F_{\theta}$, internal energy, $U_{\theta}$,
and entropy, $S_{\theta}$.
% Hence we have the free
%energy $F_{\theta}$, the internal energy $U_{\theta}$, and the
%entropy $S_{\theta}$.
The subscript $\theta$ denotes the steady
state nature of those quantities.
Due to the finite strength of stochasticity, that is, $\theta > 0$,
not all the $U_{\theta}$ is readily usable:
$F_{\theta}$ is always smaller than $U_{\theta}$.
A part of $ \theta \; S_{\theta}$ called "heat" cannot be
utilized.

It can also be verified from these definitions that if the system
consists of several non-interacting parts, $F_{\theta}$,
$U_{\theta}$, and $S_{\theta}$ are sum of those corresponding
parts. Hence, they are extensive quantities.
% No attention is paid
%here to the fine difference between additive and extensive
%properties.
The "temperature" ${\theta}$ is an intensive quantity:
it must be the same for all those parts because they are
contacting the same noise source. Therefore, we conclude that
the First Law of thermodynamics is implied by Darwinian dynamics.

The fundamental relation for the free energy, Eq.(\ref{first}), as
well as the internal energy, Eq.(\ref{i-energy}), may be expressed
in their differential forms as well. Considering an infinitesimal process
which causes changes in both the Wright evolutionary potential
function via parameter $\lambda$ and in the steady state
distribution function, the change in the internal energy according
to Eq.(\ref{i-energy}) is
\begin{eqnarray} \label{d-form1}
 d U_{\theta} & = & \int d^n {\bf q} \;
   \frac{ \phi({\bf q}; \lambda) }{\partial \lambda} \; d
   \lambda \;  \rho_{\theta}({\bf q})
    + \int d^n {\bf q} \; \phi({\bf q}; \lambda) \;
      d \rho_{\theta}({\bf q}) \nonumber \\
    & = & \mu \; d \lambda  +  \theta d S_{\theta}  \; .
\end{eqnarray}
This is the differential form for the internal energy. Here the
steady state entropy definition of Eq.(\ref{entropy}) has been
used, along with $ \int d^n {\bf q} \; d \rho_{\theta}({\bf q}) =
0$,  and
\begin{equation}
 \mu \equiv \left. \frac{ \partial U_{\theta} }{\partial \lambda } \right|
  _{\theta}  \; .
\end{equation}
Eq.(\ref{d-form1}) can be written in the usual form in
thermodynamics:
$$
  d U_{\theta} = {\bar d} W + {\bar d}Q \; .
$$

The part corresponding to the change in entropy is the "heat"
exchange: ${\bar d}Q = \theta \; d S$ and the part corresponding
to the change in the Wright evolutionary potential function is the
"work" $ {\bar d} W = \mu \; d \lambda$. The conservation of
"energy" is most clearly represented by Eq.(\ref{d-form1}). For
the free energy, following Eq.(\ref{d-form1}) and (\ref{first})
the differential form is
\begin{eqnarray} \label{d-form2}
 d F_{\theta}
  & = & d U_{\theta}  - d \theta \; S_{\theta} - \theta d S_{\theta} \nonumber \\
  & = & \mu \; d \lambda - S_{\theta} \;  d \theta  \, .
\end{eqnarray}

\subsubsection{Steady state thermodynamic definition of temperature}

Eq.(\ref{d-form1}) and (\ref{d-form2}) may be useful in some
applications. For example, the "temperature" can be found from
Eq.(\ref{d-form1}):
\begin{equation} \label{temperature}
 \theta = \left. \frac{\partial U_{\theta}}
              {\partial S_{\theta}} \right| _{\lambda} \, .
\end{equation}
There are situations in which an effective tmeprature may be needed
\cite{langer2004}. Eq.(33) may be then used to find the "temperature"
in a nonequilibrium process if it cannot be identified {\it a priori} \cite{casasvazquez2003}.

The convexity of a thermodynamic quantity is naturally
incorporated by the Boltzmann-Gibbs distribution. There is no
restriction on the size of the system. Even for a finite system,
however, phase transitions can occur, because singular behaviors
can be built into the potential function, and controlled by
external quantities.

\subsection{Second Law: maximum entropy}

\subsubsection{Second Law and Carnot cycle}

First, we remind the reader of a few important definitions. \\
A {\it reversible process} is such a process that all the relations
between quantities and parameters are defined through
the Boltzmann-Gibbs distribution, Eq.(\ref{bg}). From
Darwinian dynamics point of view, a reversible process in
reality is necessarily a slow or quasi-static process in order to
ensure the relevancy of steady state distribution
for its realization. \\
An {\it isothermal process} is a reversible process in which
"temperature" $\theta$ remains unchanged, $\theta = constant$. Do
not confuse this with thermostated processes, which are, in
general, nonequilibrium dynamical processes. \\
A {\it reversible adiabatic process} is a reversible process in
which the coupling between the system and the noise source is
switched off and the system varies in such a way that the
distribution function remains unchanged along the dynamic
trajectory at each point in phase space. The corresponding
"temperature" can be restored at any position during such a
process. This implies that the entropy remains unchanged,
$S_{\theta} = constant$. Then an irreversible adiabatic process is
one that there is no coupling between the system and the noise
environment. The system dynamics is deterministic and
conservative.

The Carnot cycle, on which the Carnot heat engine is based, is a
fundamental construction in classical thermodynamics. The Carnot
cycle consists of four reversible processes: two isothermal
processes and two reversible adiabatic processes (Fig. 2.a,b). The
efficiency $\nu$ of the Carnot heat engine is defined as the ratio
of the total net work performed over the heat absorbed at high
temperature:
\begin{equation}
 \nu \equiv \frac{ \Delta W_{total} } {\Delta Q_{12} } \; .
\end{equation}

%\begin{turnpage}
%\begin{figure}
%% \epsfig{file=fig2.eps, width=11in}
%% \epsfysize=11in
%% \epsfbox{fig22.eps}
%   \includegraphics{fig2.eps}
%%  \vspace*{1pt}
%% \caption{ \label{fig2}
%%  Carnot cycle. (a) The $\mu-\lambda$ representation.
%%  (b) The $\theta-S$ representation. In this temperature-entropy
%%  representation, the Carnot cycle is a rectangular.}
%\end{figure}
%\end{turnpage}

%\begin{turnpage}
\begin{figure}
% \epsfig{file=fig2.eps, width=8in}
% \epsfysize=9.5in
% \epsfbox{fig2.eps}
% \includegraphics{fig2.eps}
%  \vspace*{8pt}
 For the figure, please check the pubblished version: Emerging of Stochastic Dynamical Equalities and  Steady State Thermodynamics from Darwinian Dynamics, P. Ao, Communications in Theoretical Physics {\bf 49} (2008) 1073-1090. 
 \caption{ \label{fig2}
  Carnot cycle. (a) The $\mu-\lambda$ representation.
  (b) The $\theta-S$ representation. In this temperature-entropy
  representation, the Carnot cycle is a rectangular.}
\end{figure}
%\end{turnpage}

The total net work done by the system is represented by the shaded
area enclosed by the cycle. For the heat absorbed at the high
isothermal process $1 \rightarrow 2 $,
\begin{equation}
 \Delta Q_{12} = \theta_{high} \Delta S_{\theta,12}  \; .
\end{equation}
For the adiabatic process $2 \rightarrow 3$, an external
constraint represented by $\lambda$ is released (or applied),
\begin{equation}
 \Delta S_{\theta,23} = 0 {\ } ,  {\ } \Delta Q_{23} = 0  \; .
\end{equation}
For the heat absorbed (rather, released) at the low isothermal
process $3 \rightarrow 4$,
\begin{equation}
 \Delta Q_{34} = \theta_{low} \Delta S_{\theta,34} = - \Delta Q_{43}  \; .
\end{equation}
For the adiabatic process $4 \rightarrow 1$, an external
constraint is applied (released),
\begin{equation}
 \Delta S_{\theta,41} = 0 {\ } , {\ }  \Delta Q_{41} = 0 \; .
\end{equation}
Using the First Law, Eq.(\ref{first}) and the fact that the free
energy is  a state function
\begin{eqnarray}
 \Delta F_{total}
  & = & \Delta Q_{total} - \Delta W_{total}  \nonumber \\
  & = &  0  \; .
\end{eqnarray}
The minus sign in front of the total work represents that it is
the work done by the system, not to the system. The total heat
absorbed by the system is
$$
 \Delta Q_{total} =  \Delta Q_{12} + \Delta Q_{34}
   =   \Delta Q_{12} - \Delta Q_{43} = \Delta W_{total} \, .
$$
We further have
\begin{equation}
 \Delta S_{\theta,12} = \Delta S_{\theta,43} \; .
\end{equation}
From Eq.(34), (39) and (40) the Carnot heat engine efficiency is
then
\begin{eqnarray} \label{efficiency}
 \nu & = & 1 - \frac{ \Delta Q_{43}}{\Delta Q_{12} }  \nonumber \\
     & = & 1 - \frac{ \theta_{low}}{\theta_{high} }    \; ,
\end{eqnarray}
precisely the form in thermodynamics. The Second Law of
thermodynamics may be stated as that for all heat engines
operating between two temperatures, Carnot heat engine is the most
efficient. The Second Law is thus implied by Darwinian dynamics.

The beauty of Carnot heat engine is that its efficiency is
completely independent of any material details. It brings out the
most fundamental property of thermodynamics and is a direct
consequence of the Boltzmann-Gibbs distribution function and the
First Law. It reveals a property of Nature which may not be
contained in a conservative dynamics, at least it is still not
obviously to many people from Newtonian dynamics point of view
after more than 150 years of intensive studies \cite{gallavotti2004}.
On the other hand, it appears naturally implied in Darwinian
dynamics.

Having discussed various thermodynamics processes, let us return to the issue of fixing the arbitrary function in potential function discussed in Subsec. III.A, which is now directly connected to free energy. The minimum gauge condition to determine $\phi_1(\lambda)$ may be extended as
\[
  F_{\theta}(\lambda_2) - F_{\theta}(\lambda_1) 
   = \left. \left\langle W \right\rangle  \right|_{reversible} \; ,
\]    
with $ W = \int_{\lambda_1}^{\lambda_2} d{\bf q} \cdot \delta {\bf f} ({\bf q};\lambda(t))$, that is, the work done in the reversible process by the external force related to the parameter is equal to the change in free energy.
Again, it is an accepted relation in statistical thermodynamics.
It is possible that that $\phi_1(\lambda)$ determined thermodynamically may depend on temperature.

\subsubsection{Maximum entropy principle}

There are many versions of the Second Law.
%on which the
%reader is suggested to consult the books listed at the beginning
%of this section.
Here we refer to two equivalent versions from
the stability point of view, which frame the discussion in this
subsection.

{\bf Minimum free energy} statement: Given the potential
function and the temperature, the free energy achieves its lowest
possible value if the distribution is the Boltzmann-Gibbs
distribution.

{\bf Maximum entropy} statement: Given the potential function and
its average, the entropy attains its maximum value when the
distribution is the Boltzmann-Gibbs distribution.

The latter version of the Second Law is the most influential.
Its inverse statement, the so-called maximum entropy principle,
has been extensively employed in probability inference \cite{jaynes2003}
both within and beyond the physical and biological sciences
\cite{felsenstein2003, mackay2003}.

We generalize here the definitions of the entropy to include the
arbitrary time-dependent distribution in analogy to
Eq.(\ref{entropy}):
\begin{equation} \label{g-entropy}
 S(t) \equiv
  - \int d^n {\bf q} \; \rho({\bf q}, t) \; \ln \rho({\bf q}, t) \; .
\end{equation}
There are two apparent drawbacks to such a definition, however.
First, even if the evolution of the distribution function $
\rho({\bf q}, t) $ is governed by the Fokker-Planck equation,
Eq.(\ref{fp-eq}), in general the sign of the time derivative, $ d
S(t)/d t = {\dot S}(t)$, cannot be determined, whether or not it
is close to the steady state distribution. Though $ {\dot S}(t) $
might indeed be divided into an always positive part and the rest,
such a partition usually appears arbitrary. Even more problematically, in general
$S(t)$ can be either larger or smaller than $S_{\theta}$, which
makes such a definition lose its appeal in view of the
maximum entropy statement of the Second Law. We will return to
$S(t)$ later.

Nevertheless, if we take the lesson from the potential function
that only the relative value is important, we may introduce a
reference point in functional space into a general entropy definition.
One previous definition for the referenced entropy is \cite{degroot1984}
\begin{equation} \label{r-entropy}
 S_r(t) \equiv
  - \int d^n {\bf q} \; \rho({\bf q}, t) \;
   \ln \frac{ \rho({\bf q}, t) }{\rho_{\theta}({\bf q}) }
   + S_{\theta} \; .
\end{equation}
With the aid of the inequality $\ln(1+x) \leq x $ and the
normalization condition $ \int d^n {\bf q} \; \rho({\bf q}, t) =
\int d^n {\bf q} \; \rho_{\theta}({\bf q}) = 1$, it can be
verified that
\begin{eqnarray}
  S_r(t)
   & = & \int d^n {\bf q} \; \rho({\bf q}, t) \;
    \ln \left( 1 + \frac{ \rho_{\theta}({\bf q}) - \rho({\bf q}, t) }
                  {\rho({\bf q}, t) } \right)
   + S_{\theta} \; ,   \nonumber \\
   & \leq & \int d^n {\bf q} \; ( \rho_{\theta}({\bf q})
            - \rho({\bf q}), t )
           + S_{\theta} \; , \nonumber \\
   & = &  S_{\theta}  \; .
\end{eqnarray}
The equality holds when $\rho({\bf q},t) = \rho_{\theta}({\bf
q})$. This inequality is independent of the details of the
dynamics and is evidently a maximum entropy statement.
Furthermore, with the aid of the Fokker-Planck equation,
Eq.(\ref{fp-eq}), the time derivative of this referenced entropy,
$d S_r(t)/d t = {\dot S}_r(t)$ is always non-negative:
\begin{eqnarray}
 {\dot S}_r(t) \label{e-increase}
  & = & - \int d^n {\bf q} \;
    \frac{\partial \rho({\bf q}, t) }{\partial t} \;
    \ln \frac{ \rho({\bf q}, t) }{\rho_{\theta}({\bf q}) }  \nonumber \\
  & = & - \int d^n {\bf q} \;
    \left( \nabla^{\tau} [D({\bf q}) + Q({\bf q}) ] [\theta \nabla
   + \nabla \phi({\bf q}; \lambda)]  \rho({\bf q}, t) \right) \;
    \ln \frac{ \rho({\bf q}, t) }{\rho_{\theta}({\bf q}) } \nonumber \\
  & = &  \int d^n {\bf q} \;
     \left(\nabla \ln \frac{ \rho({\bf q}, t) }
                           {\rho_{\theta}({\bf q})} \right)^{\tau}
      [D({\bf q}) + Q({\bf q}) ] [\theta \nabla
   + \nabla \phi({\bf q}; \lambda)] \rho({\bf q}, t) \nonumber \\
  & = &  \int d^n {\bf q}
    \frac{1}{\theta \rho({\bf q},t) }
      \left( [\theta \nabla + \nabla \phi({\bf q}; \lambda)]
            \rho({\bf q}, t) \right)^{\tau}
    [D({\bf q}) + Q({\bf q}) ]
       [\theta \nabla+ \nabla \phi({\bf q}; \lambda)]
           \rho({\bf q}, t)  \nonumber \\
  & = &  \int d^n {\bf q} \;
     \frac{1}{\theta \rho({\bf q},t) }
     \left( [\theta \nabla + \nabla \phi({\bf q}; \lambda)]
            \rho({\bf q}, t) \right)^{\tau}
      \; D({\bf q}) \;
       [\theta \nabla+ \nabla \phi({\bf q}; \lambda)]
           \rho({\bf q}, t)  \nonumber \\
  & = &  \int d^n {\bf q} \; \frac{1}{\theta \rho({\bf q},t) }
     {\bf j}^{\tau}({\bf q}, t) \; R({\bf q}) \; {\bf j}({\bf q}, t)
      \nonumber \\
   & \geq & 0    \; .
\end{eqnarray}
Hence, this referenced entropy $S_r(t)$ has all the desired
properties for the maximum entropy statement of the Second Law.

Two important remarks are in order.
First, in the derivation reaching Eq.(45), the dynamics responsible for breaking
the detailed balance condition, the anti-symmetric matrix $Q$, does not contribute
to the change of relative entropy. Only the dissipative part of dynamics represented by
$D$ leads to the monotonic change of relative entropy.
Given the present interpretation from both trajectory and ensemble points of view,
it is clear that $Q$ is what needed for the Poisson bracket in conservative dynamics 
explored elsewhere \cite{olson2007}.
The present demonstration suggests a unified treatment for both near and far from
equilibrium dynamical processes.

Second, by the probability current density definition of
Eq.(\ref{current}) ${\bf j}$ is zero at the steady state. This is
in accordance with the understanding that at equilibrium there is
no change in (relative) entropy, that is, the entropy production
should be zero at equilibrium. Now, we have generalized this conclusion to steady
state. We note that the present probability current density may
differ from the usual probability current density definition which
may be based on Eq.(\ref{fp-generic}) and takes the form ${\bar
{\bf j}}({\bf q},t) \equiv - [\theta D \nabla - {\bar {\bf
f}}({\bf q})] \rho({\bf q}, t)$, which is not zero at the steady
state. Instead, $\nabla \cdot {\bar {\bf j}} = 0$ at the steady state.
Even with the usual definition, the zero relative entropy
production at steady state always remains valid.

Though the definition of entropy
%
%(????It is essentially a microcanonical
%definition, only good either for microcanonical ensemble, or, by
%treating it as special relative entropy of uniform potential function with
%equal steady state probability????)
of Eq.(\ref{g-entropy})
may not be appealing, a related definition of free energy is
consistent with the Second Law. We demonstrate it here. First, a
general definition for the internal energy may be:
\begin{equation} \label{g-i-energy}
 U(t) \equiv  \int d^n {\bf q} \;
   \phi({\bf q}; \lambda) \; \rho({\bf q}, t) \; .
\end{equation}
Given the distribution and the potential function, quantities
defined in Eq.(\ref{g-entropy}) and (\ref{g-i-energy}) can be
evaluated. Following the form of Eq.(\ref{f-energy}) a general
definition of free energy would be, with the "temperature"
$\theta$,
\begin{equation} \label{g-f-energy}
 F(t) \equiv U(t) - \theta \; S(t) \; .
\end{equation}
It can be verified that $F(t) \geq F_{\theta}$ and its
time derivative is always non-positive, ${\dot F}(t) \leq 0 $. So
defined time dependent free energy indeed satisfies the minimum
free energy statement of the Second Law. It differs from the
referenced entropy $S_r(t)$ by a minus sign and by a constant:
$$
  F(t) = -\theta S_r(t) + U_{\theta} \; .
$$
The generalized entropy $S(t)$ has one desired property regarding
to the adiabatic processes (either reversible or irreversible) in
that $D=0$ during the adiabatic process. Hence,
\begin{eqnarray}  \label{adiabatic}
 {\dot S}(t)
  & = & - \int d^n {\bf q} \;
    \frac{\partial \rho}{\partial t}({\bf q}, t) \;
     \ln \rho({\bf q}, t)  \nonumber \\
  & = & - \int d^n {\bf q} \;
    \left[ \nabla^{\tau} \;  Q({\bf q}) \;
      \nabla \phi({\bf q}; \lambda)  \rho({\bf q}, t) \right] \;
    \ln  \rho({\bf q}, t)  \nonumber \\
  & = & - \int d^n {\bf q} \;
    \left[(Q({\bf q})\; \nabla \phi({\bf q}; \lambda)) \cdot \nabla \right] \;
     \int^{\rho({\bf q}, t)} d \rho ' \ln  \rho '  \nonumber \\
  & = &  0 \; . {\ } {\ } {\ } (adiabatic)
\end{eqnarray}
This is the known result in conservative Newtonian dynamics that
the entropy remains unchanged. In deriving the above equation we have
used two properties: \\
1) the no-coupling to the noise environment
has been translated into the fact that the terms associated with
the diffusion matrix $D$ and "temperature" $\theta$ are set to be
zero in Eq.(\ref{fp-eq}), because they are related to the noise
source which is decoupled during an adiabatic process; and \\
2) the incompressible condition of $ \nabla \cdot [
Q({\bf q}) \; \nabla \phi({\bf q}; \lambda) ] = 0 $, the Liouville theorem,
which is typically satisfied in Newtonian dynamics. \\
In this conservative case, it can be verified that ${\dot S}_r(t)
= 0$, too, for any adiabatic process. Nevertheless, it is possible
that while the dynamics is conservative, $R=0$, and even
satisfying the Jacob identity in classical mechanics, the
Liouville theorem can be violated. Hence, in this case the general
entropy $S(t)$ is not a constant during the dynamical process, and
the general free energy $F(t)$ and the referenced entropy $S_r(t)$
are not well defined either, because there is no "temperature" to
define a steady state distribution. But the "energy" can be
conserved during such a process \cite{olson2007}.

\subsubsection{Connection to information theories}

It may be worthwhile to define another referenced entropy
$S_{r2}(t)$  which approaches the steady state entropy
$S_{\theta}$ from above. It's form is simple:
\begin{equation}
 S_{r2}(t) \equiv - \int d^n {\bf q} \;
    \rho_{\theta}({\bf q}) \; \ln \rho({\bf q}, t) \; .
\end{equation}
It can be verified that $S_{r2}(t) \geq S_{\theta}$ and ${\dot
S}_{r2}(t) \leq 0 $.

The relation defined by Eq.(\ref{r-entropy}) is in the same form
as the relative information in information theories. Ref. \cite{cover2006} 
contains discussions of many other useful
inequalities. A rather complete coverage of information theory can
be found in Ref. \cite{mackay2003}, and some current discussions of its
connection to thermodynamics can be found in Ref. \cite{bialek2001}
and \cite{garbaczewski2006}.

\subsection{Third Law: unattainability of zero "temperature"}

Now we consider the behavior near zero "temperature", $\theta
\rightarrow 0$.  To be specific, we assume the system is dominated
by a stable fixed point, taking as ${\bf q} = 0$.
As suggested by the Boltzmann-Gibbs
distribution, Eq.(\ref{bg}), only the regime of phase space near
this stable fixed point will be important. Hence the Wright
evolutionary potential function can be expanded around this point:
\begin{equation}
 \phi({\bf q}; \lambda) = \phi(0; \lambda) + \frac{1}{2}
  \sum_{j=1}^{n} k_j(\lambda) q_j^2  {\ } .
\end{equation}
Here we have also assumed that the number of independent modes is
the same as the dimension of the phase space, though it may not
necessary be so. This assumption will not affect our conclusion
below. Those independent modes are represented by $q_j$ without
loss of generality.  The "spring coefficients" $ \{ k_j \}$ are
functions of external parameters represented  by $\lambda$.

The partition function according to Eq.(\ref{p-function}) can be
readily evaluated in this situation:
\begin{equation}
 {\cal Z}_{\theta} = e^{-\beta \phi(0; \lambda)}  \; \prod_{j=1}^{n}
   \sqrt{\frac{2\pi \theta }{k_{j}} } {\  } .
\end{equation}
So, the entropy according according to Eq.(\ref{entropy}), is:
\begin{equation}
 S_{\theta} =  n  \left[ \theta -  \frac{1}{2} \ln\theta \right]
  + \frac{1}{2} \sum_{j}^{n} \frac{k_j}{2\pi}\, .
\end{equation}
The first term does not depend on external parameters, but the
second term does.  This suggests that the entropy depends on
the specific control process to achieve low temperature:
different processes would lead to different sets of $\{k_j\}$,
hence a difference between entropies at low temperatures.
The Third Law states that in the limit of zero temperature the
difference in entropy between different processes is zero.
Thus, Darwinian dynamics as formulated in the present paper
does not imply the Third Law.

One should not be surprised by the above conclusion, because
Darwinian dynamics here is essentially "classical".
The same conclusion could also be reached from classical Newtonian dynamics.
This means that within Darwinian dynamics one could easily conceive the
zero "temperature" limit without any logic inconsistency.

With quantum mechanics, the agreement to the Third Law is found and a
stronger conclusion is reached: Not only the difference in
entropy should be zero, the entropy itself is zero at zero
temperature.
%
%(???? connection to the no-cloning theorem????)
%
We may conclude that, in general, completely neglecting the noise
is not a viable choice: The temperature cannot be zero.
When the noise is small enough, new phenomena may occur \cite{landsberg1998}.
Phrasing differently, there appears to exist a bottom near which
there is something. It should be pointed out that in the present
formulation of Darwinian dynamics, particularly the
Eq.(\ref{normal}), the existence of an anti-symmetric matrix
suggests a natural route to define a Poisson bracket. Therefore,
it is possible to extend Darwinian dynamics into the quantum
regime by following the usual canonical quantization procedure, possibly following
the suggestions from dissipative quantum dynamics \cite{ao1991, leggett1992,
griffiths2002}. Studies show that the Third law can be regained in this way
\cite{hanggi2006, ford2007}.

%
%  Comment: T = \infty not attanable classically, because of infintite degrees
% of freedoms. But T = 0 is in principle attenable.
% Quantum mechanics is other way around: because of finite degrees of freedom, 
% T =  0 would not be accessible but T = \infty can.
%

\subsection{Two inferences}

To summarize, in this section we have shown that except for the Third
Law, all other Laws of thermodynamics follow from Darwinian
dynamics. The concern \cite{sekimoto1998} as to which
stochastic integration method, Ito, Stratonovitch-Fisk,
Hanggi-Klimontovich,
or others, is consistent with the Second Law is resolved: Any of
them can be decomposed into three parts: the conservative dynamics represented by 
the antisymmetric matrix $Q$, and nonconservative dynamics represented by nonnegative symmetric matrix $D$, and the potential function $\phi$. Thus, any of them is consistent with the Second Law. We also
note that based on thermodynamic relations, the fundamental
relation of Eq.(\ref{first}), the conservation of energy of
Eq.(\ref{d-form1}), the universal heat engine efficiency of
Eq.(\ref{efficiency}), supplemented by the additive nature of extensive
quantities and the temperature of Eq.(\ref{temperature}), the
Boltzmann-Gibbs distribution is implied. In this way
statistical mechanics and thermodynamics are equivalent.

Thermodynamics deals with steady state properties. The key
property is determined by the Boltzmann-Gibbs distribution of
Eq.(\ref{bg}) which only depends on the Wright evolutionary
potential function $ \phi $ and the "temperature" $ \theta $ of Darwinian dynamics.
The rest relations are determined by the various symmetries of the
system. No dynamical information can be inferred from them. This
feature has been noticed in the literature \cite{zeh2007}. In
particular, there is no way to recover the information on two
quantities determining the local time scales, the friction matrix
$ R $ and the transverse matrix $ T $, from thermodynamics. In
this sense "time" is lost in thermodynamics.
Thus, thermodynamics contains no direction of time and hence is consistent with the time-reversal conservative Newtonian dynamics.

\section{Stochastic Dynamical Equalities }

We have explored the steady state consequences of Darwinian
dynamics in statistical mechanics and in thermodynamics. In this
section we explore its general dynamical consequences. Two types
of recently discovered dynamical equalities will be discussed: one
based on the Feynman-Kac formula and other a generalization of
Einstein relation. For a background on path integral formulation,
Feynman's lucid exposition is highly recommended \cite{feynman1972}.

\subsection{Feynman-Kac formula}

Previous discussions demonstrate that the Boltzmann-Gibbs
distribution plays a dominant role. It is natural to work in a
representation in which Boltzmann-Gibbs distribution appears in a
most straightforward manner, or, as close as possible. The
standard approach in this spirit is as follows. First, choose the
dominant part of the evolution operator $L$. The remaining part is
denoted as $\delta L$. In this subsection a general methodology to
carry out this procedure is summarized.

The Fokker-Planck equation, Eq.(\ref{fp-eq}), can be rewritten as
\begin{equation} \label{fp-eq3}
 \frac{ \partial}{\partial t} \rho({\bf q},t)
  = L (\nabla,{\bf q}; \lambda) \rho({\bf q},t) \, ,
\end{equation}
with $ L = \nabla^{\tau} [D({\bf q}) + Q({\bf q})]
   [\theta  \nabla + \nabla \phi({\bf q}) ]$.
It's solution can be expressed in various ways. The most
suggestive form in the present context is that given by Feynman's
path integral \cite{feynman1972}:
If at time $t'$ the system is at ${\bf q}'$, the probability for
system at time $t$ and at ${\bf q}$ is given by summation of all
trajectories allowed by Eq.(\ref{normal}) connection those two
points:
\begin{equation}
 \pi({\bf q},t; {\bf q}', t')
  = \left. \left. \sum_{trajectories}
   \right\{ {\bf q}(t) = {\bf q};
            {\bf q}(t') = {\bf q}' \right\} \; .
\end{equation}
In terms of the summation over the trajectories, the solution to
Eq.(\ref{fp-eq3}) (and Eq.(\ref{fp-eq})) may be expressed as
\begin{eqnarray} \label{trajectory}
 \rho({\bf q},t)
  & = & \int d^n {\bf q}' \, \pi({\bf q},t; {\bf q}', t') \,
   \rho({\bf q}', t=0)  \nonumber \\
  & \equiv & \left. \left\langle
     \delta({\bf q}(t)- {\bf q}) \right\rangle \right| _{trajectory} \; .
\end{eqnarray}
The delta function $\delta({\bf q}(t)-{\bf q})$ is used to
specify the end point explicitly. There is an additional summation over
initial points, ${\bf q}'$, weighted by the initial distribution
function, $\rho({\bf q}',t=0)$.

Now, considering that the system is perturbed by $\delta L({\bf
q}; \lambda)$, represented, for example, by a change in control
parameter $\lambda$. The new evolution equation is
\begin{equation}
 \frac{ \partial}{\partial t} \rho_{new}({\bf q},t)
  = [ L(\nabla,{\bf q}; \lambda) + \delta L(\nabla, {\bf q}; \lambda) ] 
       \rho_{new}({\bf q},t) \, .
\end{equation}
The perturbation may act as a source or sink for the probability
distribution. The probability is no longer conserved: in general
$\int d{\bf q} \; \rho_{new}({\bf q},t) \neq \int d{\bf q} \;
\rho_{new}({\bf q},t=0) $. According to the Feynman-Kac formula
\cite{delmoral2004},
the solution to this new equation can be expressed as
\begin{equation} \label{feynmankac}
 \rho_{new}({\bf q},t)
   = \left. \left\langle \delta({\bf q}(t) - {\bf q}) \;
     e^{\int_0^t dt'\; \delta L(\dot{\bf q}(t'), {\bf q}(t'); \lambda ) }
     \right\rangle \right| _{trajectory} \; ,
\end{equation}
with $ \rho_{new}({\bf q}',t=0) = \rho({\bf q}',t=0)$ and the
trajectories following the dynamics of Eq.(\ref{normal}), the same
as that in Eq.(\ref{trajectory}). 
Thus, the evolution of the distribution function under new dynamics 
can be expressed by the evolution in the original
dynamics. The corresponding procedure in quantum mechanics is that
in the "interaction picture" \cite{schiff1968}.
Eq.(\ref{feynmankac}) is a powerful equality. Various dynamical
equalities can be obtained starting from Eq.(\ref{feynmankac}).
Indeed, its direct and indirect consequences have been extensively
explored \cite{bochkov1977, evans2002}.

\subsection{Free energy difference in dynamical processes}

\subsubsection{Jarzynski equality}

We have noticed the special role played by the Boltzmann-Gibbs
distribution, Eq.(\ref{bg}). In particular, it is independent of
the friction and transverse matrices $R,T$. Evidently the
instantaneous Boltzmann-Gibbs distribution with $\lambda =
\lambda(t)$ is
\begin{equation} \label{i-bg}
 \rho_{\theta}({\bf q}; \lambda(t))
  = \frac{ e^{ - \beta \phi({\bf q}; \lambda(t)) } }
         {{\cal Z}_{\theta}(\lambda (0))} \, .
\end{equation}
Here we have explicitly indicated that the parameter $\lambda$ is
time-dependent. This distribution function is no longer the
solution of the Fokker-Planck equation of Eq.(\ref{fp-eq}), however.
There will be transitions out of this instantaneous Boltzmann-Gibbs
distribution function due to the time-dependence of the parameter
$\lambda$. While such transitions may be hard to conceive of in
classical mechanics, they can be easily rationalized in quantum
mechanics, because of the discreteness of states.
One such well studied model is the dissipative Landau-Zener
transition \cite{kayanuma1984, ao1989, ao1991}.

An interesting question is that whether the transitions can be
"reversed" such that the instantaneous distribution is indeed an
explicit solution for another but closely related evolution
equation. This means that the original Fokker-Planck equation has
to be modified in a special way to become a new equation. Indeed,
this modified evolution equation can be found for any function
${\bar \rho}({\bf q},t)$, which reads,
\begin{equation} \label{new-eq}
 \frac{ \partial}{\partial t} \rho_{new}({\bf q},t)
  = \left[ L(\nabla,{\bf q},t) - \frac{1}{{\bar \rho}({\bf q},t)}
      ( L(\nabla,{\bf q},t) {\bar \rho}({\bf q},t) )
     + \left( \frac{\partial  \ln | {\bar \rho}({\bf q}, t)  | }
         {\partial t} \right) \right]
     \rho_{new}({\bf q},t) \, .
\end{equation}
It can be verified  $ \rho_{new}({\bf q}, t) = {\bar \rho}({\bf
q},t)$ is a solution of above equation. Treating
$$
 \delta L =
  - \frac{1}{{\bar \rho}({\bf q},t)}
    L(\nabla,{\bf q},t) {\bar \rho}({\bf q},t)
  + \frac{\partial \ln|{\bar \rho}({\bf q}, t)| }
         {\partial t}
$$
and the Feynman-Kac formula Eq.(\ref{feynmankac}) may be applied.
The analogous procedure has been well studied for transitions
during adiabatic processes in interaction picture of quantum
mechanics \cite{schiff1968, ao1989, ao1991}
and of statistical mechanics \cite{feynman1972}.

Now, let $\bar \rho$ be the instantaneous Boltzmann-Gibbs
distribution of Eq.(\ref{i-bg}): ${\bar \rho} = \rho_{\theta}({\bf
q}; \lambda(t))$. We have
$$
  \delta L = -\beta  {\dot \lambda} \frac{\partial \phi ({\bf q}; \lambda )}
                 {\partial \lambda} \; .
$$
Eq.(\ref{new-eq}) can be solved by summing over all trajectories
using the Feynman-Kac formula, Eq.(\ref{feynmankac}). At the same
time, we know the instantaneous Boltzmann-Gibbs distribution of
Eq.(\ref{i-bg}) is its solution. Since these two are solutions to
the same equation, we have following equality
\begin{equation} \label{feynman-kac2}
 \frac{ e^{ - \beta \phi({\bf q}; \lambda(t)) } }
      {\int d {\bf q} \, e^{ - \beta \phi({\bf q}; \lambda(0)) }}
  =  \left.  \left\langle
     \delta({\bf q} - {\bf q}(t) ) \;
    e^{ - \beta \int_{0}^{t} d t' \; {\dot \lambda}(t')
     \frac{ \partial \phi({\bf q}(t'); \lambda(t')) }
          {\partial \lambda} } \right\rangle  \right|_{trajectory} \, .
\end{equation}
Following Ref. \cite{jarzynski1997}
we define the dynamical work along a trajectory as
\begin{equation}
 W_{\lambda} = \int _{0}^{t} d t' \; {\dot \lambda}(t')
     \frac{ \partial \phi({\bf q}(t'); \lambda(t')) }
          {\partial \lambda}  \, .
\end{equation}
The equality between the free energy difference $\Delta F_{\theta}
= F_{\theta}(t) - F_{\theta}(0) $ and the dynamical work $W_{\lambda} $ is,
then, after summation over all final points of the trajectories in
Eq.(\ref{feynman-kac2}),
\begin{equation} \label{jarzynski}
 e^{ -\beta \Delta F_{\theta} } =  \left. \left\langle
    e^{ - \beta W_{\lambda} } \right\rangle  \right|_{trajactory}  \, .
\end{equation}
This elegant equality connects the steady state quantities $\Delta
F_{\theta}$ to the work done in a dynamical process. 
Such parametrized form was first discovered by Jarzynski \cite{jarzynski1997}.
It should be emphasized that there is no assumption of steady
state state at time $t$ for the system governed by
Eq.(\ref{fp-eq}). In fact, it is known, for example, in the case
of the Landau-Zener transition that it is not steady state \cite{ao1989, ao1991}.
This equality has been discussed and extended by various authors
from various perspectives \cite{crooks1999, hummer2001, hummer2005, hatano2001, mukamel2003,
park2004, chernyak2005, qian2005b}.
The connection of this equality to the Feynman-Kac formula was
first explicitly pointed out in Ref. \cite{hummer2001}.
There have also been experimental verifications of this equality
\cite{collin2005}. This type of equalities was reviewed recently \cite{kurchan2007}.

Three remarks may be made here.
First, the derivation of Jarzynski equality presented here is valid
both with and without detailed balance condition,
with both additive and multiplicative noises. It has only one result.
Second, for Jarzinskii equality, neither $D$ nor $Q$ enter into the equality, 
while the dynamics are obviously determined by those matrices.
Third, Feynman-Kac formula may be used to generate more dynamical equalities,
as already noted. 

In the light of those observations, we may infer three immediate but somewhat surprising physical results.
First, the "temperature" can be time dependent, too. 
Thus, a work equality for "temperature" can be established by explicitly going through the procedure, hence extends the work relation to a different dynamical domain.
Second, the validity of the demonstration of Jarzynski equality does not depend on the details 
of the operator $L$ in Eq.(53), as long as the steady state exists. This would suggest that colored noises 
can be entered into Eq.(53). In fact, we already know such examples \cite{ao1989, ao1991, leggett1992}.
The dynamical equation equation as expressed by Eq.(4) also allows a straightforward extension to colored noises. 
Third, given the potential function determined by a reversible process discussed in Sec. IV.C.1 and that using Feynman-Kac formulae we have \cite{bochkov1977, schurr2003, astumian2006} 
\begin{eqnarray}
   &   &  \left. \left\langle \exp\{ -( W - W_{\lambda} ) \}
                 \right\rangle \right|_{trajectory} \nonumber \\ 
   & = & \left. \left\langle \exp\left\{ - \int d{\bf q}\cdot \delta {\bf f}({\bf q}(t'); \lambda(t_1))
    + \int _{0}^{t} d t' \; {\dot \lambda}(t') \partial_{\lambda}
      \phi({\bf q}(t'); \lambda(t')) \right\} \right\rangle \right|_{trajectory} \nonumber \\ 
   & = & 1   \, , \nonumber 
\end{eqnarray} 
Jarzynski equality may be used to check consistency in our understanding of related dynamical quantities. For example, a discrepancy may indicate possibly a missing term in the potential function $\phi$.

\subsubsection{Microcanonical and canonical ensembles}

The Jarzynski equality places the Boltzmann-Gibbs distribution
hence the canonical ensemble in the central position. They are
simply natural consequences from Darwinian dynamics. However,
if we start from conservative, Newtonian dynamics, the
appropriate ensemble is the microcanonical ensemble. Any
distribution function which is a function of the potential
function or Hamiltonian would be the solution of the Liouville
equation. From this point of view the Boltzmann-Gibbs distribution
and the associated temperature appear arbitrary: It is just one
among infinite possibilities. This concern has been raised in
the literature \cite{cohen2005b}
regarding to the generality of the equality of
Eq.(\ref{jarzynski}). No satisfactory treatment of this concern
within Newtonian dynamics has been given. Rather, it has been an
"experimental attitude": If one does this and makes sure the
procedure is correct one gets that, and it works. Darwinian
dynamics, however, provides an {\it a priori} reason to fully
justify the use of the Boltzmann-Gibbs distribution in the
derivation of the Jarzynski equality.

\subsection{Generalized Einstein relation}

In deriving the Boltzmann-Gibbs distribution from Darwinian
dynamics, Eq.(\ref{einstein}):
$$
  [R({\bf q}) + T({\bf q}) ] \, D({\bf q}) \,
     [ R({\bf q}) - T({\bf q}) ] = R({\bf q})  \; ,
$$
has been used.
This general and simple dynamical equality was termed the
generalized of Einstein relation \cite{ao2004}.
If the detailed balance condition is satisfied, that is, if $T = 0$ or $Q = 0$,
the above relation reduces to $ R D = 1$, which was discovered a century ago
by Einstein \cite{einstein1905},
% \cite{einstein}
and since then has been known as Einstein relation. Variants of
Einstein relation in different settings were obtained earlier and
independently by Nernst \cite{nernst1888},
by Planck \cite{planck1890},
by Townsend \cite{townsend1899},
and by Sutherland \cite{sutherland1905}.
Similar to the dynamical equality exemplified by Jarzynski
equality, the generalized Einstein relation is a consequence of
the Boltzmann-Gibbs distribution and the canonical ensemble
embedded in Darwinian dynamics.

Experimentally, all those quantities in Eq.(\ref{einstein}) can be
measured independently. Hence, this generalized Einstein relation
should be subject to experimental tests in the absence of detailed
balance, that is, when the antisymmetric matrix $ T $ does not
vanish. While in evolutionary processes in biology the data can be
organized by the present dynamical structure \cite{li1977}, the
parameters are typically fixed by Nature. We need situations where
all these elements, $R,T,\phi,\theta$, are accessible to
experimental control.

For simplicity, we consider a nonequilibrium situation realizable
with current technology as an illustration: a charged nanoparticle
or macromolecule, an electron or a proton, with charge denoted by
$e$, in the presence a strong, uniform magnetic field $B$ and
immersed in a viscous liquid with friction coefficient $\eta$.
Indeed, similar situation has already been considered
experimentally \cite{blickle2007}.
Here we restrict our attention to the two
dimensional case ($n=2$). The corresponding Darwinian dynamical
equation of Eq.(\ref{normal}) in this case is the Langevin
equation with the Lorentz force for a
"massless" charged particle \cite{ashcroft1976}:
\begin{equation}
 \eta \dot {\bf q} + \frac{e}{c} B \hat{z} \times \dot {\bf q}
   = - \nabla \phi({\bf q}) + N_{II} {\bf \xi}(t)
\end{equation}
The friction matrix is
\begin{equation}
 R = \eta \left( \begin{array}{ll}
           1 & 0 \\
           0 & 1 \end{array}
      \right)
\end{equation}
The transverse matrix is
\begin{equation}
 T = \frac{e}{c} B  \left( \begin{array}{ll}
           0 & 1 \\
           -1 & 0 \end{array}
      \right)
\end{equation}
and the "temperature" is $\theta = k_B T_{BG}$, with the Boltzmann
constant $k_B$ and the thermal equilibrium temperature $T_{BG}$.
If the system is out of thermal equilibrium, the effective
temperature, such as defined by Eq.(\ref{temperature}), should be
used. This is a physically realizable situation in two
dimensional electrons extensively discussed recently \cite{rodrigues2006}.
The corresponding Fokker-Planck equation, following Eq.(\ref{fp-eq}), is
\begin{equation}
 {\partial \rho({\bf q},t) \over \partial t}
  = \nabla [ D \theta \; \nabla
   + [D + Q ] \nabla \phi({\bf q})] \rho({\bf q},t) \; .
\end{equation}
This is precisely a diffusion equation with diffusion matrix
$D$. Both $D$ and $Q$ can be obtained from the generalized Einstein
relation, Eq.(\ref{einstein}):
\begin{equation}
 D = \frac{\eta}{\eta^2 + \left( \frac{e}{c} B \right)^2 }
      \left( \begin{array}{ll}
           1 & 0 \\
           0 & 1 \end{array}
      \right)
\end{equation}
\begin{equation}
 Q = \frac{ \frac{e}{c} B }{\eta^2 + \left( \frac{e}{c} B \right)^2 }
      \left( \begin{array}{ll}
           0 & -1 \\
           1 & 0 \end{array}
      \right)
\end{equation}
In a typical situation, though all quantities can be measured
experimentally, the friction coefficient is likely to be less
sensitive to the magnetic field. Then one may need to focus
experimentally on the diffusion in the presence of the magnetic
field without any potential field. In this case the evolution of
the distribution is governed by the standard diffusion equation:
\begin{equation} \label{diffusion}
 {\partial \rho({\bf q},t) \over \partial t}
  =    \theta d_B \; \nabla^2 \rho({\bf q},t) \; ,
\end{equation}
with
$$
 d_B = \frac{ \eta }{ \left[\eta^2 + \left( \frac{e}{c} B \right)^2
       \right] } \, .
$$ 
The solution to Eq.(\ref{diffusion}) with $\rho({\bf q},t=0 = \delta {\bf q}(t=0) - {\bf q}$ is standard (two dimension, $n=2$):
$$
   \rho({\bf q},t) = \frac{ 1 }{2\pi \; t }
    \exp \left\{ - \frac{ {\bf q }^2 }{2d_B \theta \; t } \right\}
$$
Averaging over trajectories governed by Eq.(\ref{diffusion}), $
\left. \langle {\bf q}(t) - {\bf q}(t=0) \rangle
\right|_{trajectory} = 0 $ and
$$
%\begin{equation}
 \left.  \left\langle ({\bf q}(t) - {\bf q}(t=0) )^2 \right\rangle
  \right|_{trajectory} =  4 d_B \theta \; t  \; .
%\end{equation}
$$
An experimental system, for example, may be that of injection of
electrons into a semiconductor, wherein one measures their
diffusion in the presence of a magnetic field. Every quantity in
the generalized Einstein relation of Eq.(\ref{einstein}) can be
measured and controlled experimentally.
%Such experiments may has already been
%done.
%
% (????).
%
Another experimental system may be on ionized hydrogen or
deuterium. For charged macromolecules and nano-particles, the
friction coefficient may be too large to allow a measurable
magnetic field effect accessible by current magnets. As a
numerical example, for the zero magnetic field diffusion constant
of $d_{B=0} k_B T_{BG} \sim 10^4 \; cm^2/sec$, which implies
diffusing of about $100 cm$ in 1 second, the friction coefficient
is $\eta = 1/d_{B=0} \sim 4 \times 10^{-16} dyne/(cm/sec) $ at
temperature $T_{BG} = 300 K$.
%
%  The typical ionic diffusion in a liquid is order of
%   $ 2 \times 10^{-5} \; cm^2/sec$,
%
Assuming one net
electron charge, for magnetic field $ B= 1 \; Telsa$ , we have $eB
/ c \sim 1.6 \times 10^{-16} dyne/(cm/sec) $, comparable to the
friction coefficient.
% They are experimentally accessible(????).

%\section{Discussions}

%\subsection{literature}

%\subsection{speculative discussions}

% EH Lieb: no single violation of the Second Law so far!

%ensemble vs assemble

%existence vs realization

%probability distribution vs density

\section{Outlook}

In the present paper we have presented statistical mechanics and
steady state thermodynamics as natural consequences of Darwinian
dynamics. Two types of general stochastic dynamical equalities
have been explored. Both can be directly tested experimentally.
Everything appears fully consistent except for one issue. The
point of view in physics has been that we should start from
conservative dynamics, not Darwinian dynamics. This view has
indeed strong experimental and historical supports during past 150
years. It is still the subject of current research \cite{lebowitz1999,
gallavotti2004, cohen2005a, goldstein2006}.
The troubling issue may be expressed by an attempt to answer the following
question. The natural consequence of conservative dynamics is the
micro-canonical ensemble, from which the canonical ensemble just
appears to be one of its infinite possibilities. How and why does
Nature choose the canonical ensemble and the Second Law? There
does not seem to be a consensus yet on the answer.

The difficulty in reaching the Second Law from conservative
dynamics in nonequilibrium setting may encourage us to consider
Darwinian dynamics. There is, however, a more
compelling reason: Darwinian dynamics is the most fundamental
and successful dynamical theory in biological sciences.
Furthermore, as we have demonstrated above, the Second Law
and other nonequilibrium properties follow naturally from it.
Logically it provides a simple starting point.
It must contain a large element of physical truth.

Conservative dynamics and Darwinian dynamics appear to
occupy the two opposite ends of the theoretical description of
Nature. Both have been extremely successful. In many respects they
appear to be complementary to each other.
For example, it was noticed that Darwinian dynamics and
Newtonian dynamics can be derived from each other under
appropriate conditions \cite{ao2005}.
What would be the implication of this mutual reduction?
Are there hidden reasons for such complementarity?
Hints to answers for such questions are perhaps already contained
in the discussions of "more is different" \cite{anderson1972, laughlin2000},
of the immensity of functional space \cite{elsasser1998},
%
% on immensity:
% "Only two things are infinite, the universe and human stupidity,
% and I'm not sure about the former."  Albert Einstein
%
of the macroscopic quantum effect \cite{leggett1992},
and of universe vs multiverse \cite{carr2007}.
The formulation and analysis in this paper may provide insights into those 
fundamental relationships and a stimulus for further studies. 

%
% ensemble vs assemble
% realization of thermodynamics
%  Bose-Einstein, Dirac-Fermi,
%  q-statitics
%

{\ }

{\ }

{\bf Acknowledgements}. I thank
M. Dykman, G.L. Eyink, J. Felsenstein, D. Galas, P. Gaspard, P. Hanggi, B.L. Hao, C.
Jarzinski, M. Kardar, C. Kwon, P.A. Lee, A.J. Legggett, H. Qian, J. Rehr, L.S. Schulman, D.J.
Thouless, R.E. Ulanowicz, J. Wang, S.T. Yau, L. Yin, L. Yu, H.D. Zeh,  
X.M. Zhu for constructive comments and critical discussions at various
stages of this work. I also thank D. Galas for a critical reading
of the manuscript.
This work was supported in part by USA NIH grant under K25-HG002894. \\
Note on reference. There is a vast body of work on statistical
mechanics, thermodynamics, evolution, and related topics. No single paper can
do justice to the relevant literature. Admittedly incomplete, it
is my hope that a useful fraction of literature has been covered
and that a spirit of current research activities has been
captured. In addition, this work is a critical discussion of two
fundamental fields based on an emerging dynamical formulation.
Biases and prejudices are unavoidable. I apologize to those whose
important works are not mentioned here, likely the result of my
oversight. I would appreciate the reader's effort very much to
bring her/his and/or other's important works to my attention (E-mail:
 aoping@u.washington.edu).

\end{document}